\documentclass[10pt]{revtex4}
\usepackage{amsmath,amssymb}
\usepackage{graphicx,epstopdf}
\usepackage[usenames,dvipsnames,svgnames]{xcolor}
\usepackage{hyperref}
\usepackage{array}
\usepackage{multirow}
\usepackage{blindtext}
\usepackage{relsize}

\begin{document}
\title{Some specific wormhole solutions in $f(R)$-modified gravity theory}
\author{Bikram Ghosh$^1$\footnote {bikramghosh13@gmail.com}}
\author{Saugata Mitra$^1$\footnote {saugatamitra20@gmail.com}}
\author{Subenoy Chakraborty$^2$\footnote {schakraborty.math@gmail.com}}
\affiliation{$^1$Department of Mathematics, Ramakrishna Mission Vidyamandira, Howrah-711202, West Bengal, India\\
	$^2$ Department of Mathematics, Jadavpur University, Kolkata-700032, West Bengal, India}


\begin{abstract}
	The paper deals with the static spherically symmetric wormhole solutions in $f(R)$-modified gravity theory with anisotropic matter field and for some particular choices for the shape functions. The present work may be considered as an extension of the general formalism in \cite{r15} for finding wormhole solutions. For isotropic matter distribution it has been shown that wormhole solutions are possible for zero tidal force and it modifies the claim in \cite{r17}. Finally energy conditions are examined and it is found that all energy conditions are satisfied in a particular domain with a particular choice of the shape function.	
	
\end{abstract}
\maketitle

\section{Introduction:} In astrophysics, there is a popular concept, namely interstellar travel which is commonly described by traversable wormhole \cite{r1}. The basic property for the traversability of such wormhole is that they do not possess any horizon, it has throat (having minimum surface area) with a flare-out condition \cite{r1.1} as a consequence there is a violation of null energy condition (NEC) by the matter stress tensor at the throat. In spherical polar coordinate in the background of spherically symmetric, spacetime the traversable wormhole is described by the standard line element \cite{r2}
\begin{equation} \label{eq1} ds^2=-e^{2\Phi(r)}dt^2+\left[1-\frac{b(r)}{r}\right]^{-1}dr^2+r^2(d\theta^2+sin^2\theta d\phi^2),
 \end{equation}
where $\Phi(r)$ is the gravitational redshift function and $b(r)$ is the shape function indicative of the shape of the wormhole throat. The radial co-ordinate `$r$' decreases from infinity to a minimum value $r_0$, where $b(r_0)=r_0$ and then increases from $r_0$ back to infinity. Here, in order for the wormhole to be traversable, one should avoid the formation of an event horizon \cite{r3} which are identified as the surface
s with infinite redshift {\it i.e.} $e^{2\Phi}\longrightarrow0$ at the horizon, so that $\Phi(r)$ must be finite everywhere for traversability. Also a fundamental property of wormholes is the flaring out of the throat, which is translated by the condition $(b-b^\prime r)/b^2>0$ at $r=r_0$ \cite{r4}. At the throat, the condition $b^\prime(r_0)<1$ is imposed in order to have wormhole solutions \cite{r4}-\cite{r5}. It is well known that wormholes in classical general relativity require a violation of the null energy condition, and the corresponding matter is termed as ``exotic matter" \cite{r6}. Such matter must be confined to a very narrow band around the throat.
\par 
 However, recently it has been shown in the context of modified theories of gravity that the matter threading the wormhole may satisfy the energy conditions but the effective stress-energy tensor \cite{r7} involving higher order derivatives is responsible for the violation of the NEC. There are several wormhole solutions in modified gravity theories, for examples, in Brane-world gravity wormhole solutions are obtained by various technique \cite{b1}--\cite{b5}, in Einstein-Gauss-Bonnet gravity, traversable wormhole solutions \cite{n1}--\cite{n2} satisfy weak energy conditions, in $f(R, T)$ gravity theory, thin-shell wormhole solutions \cite{n3}--\cite{n7} are obtained for suitable choice of `$f$' and it is found that these solutions satisfy both null and weak energy conditions \cite{n8}--\cite{n9}, in Scalar-Tensor gravity traversable and spinning wormholes are found in the literature \cite{n10}--\cite{n11} while in Lovelock gravity theory various wormholes ({\it i.e. }Lorentzian, Cosmological and Classical) are obtained \cite{n12}--\cite{n14}.
 \par
 Over the past decade, $f(R)$ gravity theory has been extensively studied as one of the simplest modifications to Einstein-Hilbert action. In this modified gravity theory, wormhole geometries and some exact solutions are obtained \cite{n15} by various choices of the function `$f$'. It is found that for the above wormhole solutions, not only the null energy condition is satisfied, but also weak and dominant energy conditions are also satisfied \cite{n16}--\cite{n17}. Also, both static and evolving wormhole solutions are obtained in recent past \cite{n18}--\cite{n19}. In most of these works, viability bounds have been explored for energy conditions considering both imperfect as well as perfect fluid models. Further, wormholes solutions have been constructed in some cases through cut and paste technique.
\par
In this work, $f(R)$-modified gravity theory has been considered for obtaining wormhole solutions both for anisotropic and isotropic matter fields. For anisotropic matter field, wormhole solutions are obtained using the general mechanism in \cite{r15} for Einstein gravity. Also, energy conditions are discussed in each case. For isotropic matter field, wormhole solutions so obtained are examined whether they are in favour or against the conclusion in \cite{r17}. The paper is organized as follows: Basic equations in $f(R)$-modified gravity theory have been presented in section \ref{secii}. Section \ref{seciii} deals with wormhole solutions with anisotropic matter field and energy conditions are examined. In section \ref{seciv}, wormhole solutions are constructed for isotropic matter field, and in section \ref{secv}, embedding diagrams of wormholes are shown. Finally, summary and concluding remarks are given in section \ref{secvi}.
 
\section{Basic equations in $f(R)$-modified gravity theory}\label{secii}
In $f(R)$-modified gravity, Ricci scalar $R$ is replaced by an arbitrary function $f(R)$ in the Einstein-Hilbert action \cite{r8} $$S_{EH}=\int\sqrt{-g}Rd^4x$$ becomes :$$S_{f(R)}=\int\sqrt{-g}f(R)d^4x.$$ In $f(R)$ gravity model, action is given by \cite{r9}
\begin{equation}\label{eq2}
S=\frac{1}{2k^2}\int d^4x\sqrt{-g}f(R)+S_M(g^{\mu\nu},\psi)
\end{equation} where $k^2=8\pi G$, and for notational simplicity we consider $k^2=1$ in this work. $S_M(g^{\mu\nu},\psi)$ is the matter action, defined as $S_M=\int d^4x\sqrt{-g}\mathcal{L}_M(g_{\mu\nu},\psi)$, where $\mathcal{L}_M$ is the matter Lagrangian density, where matter is assumed to be minimally coupled to gravity and $\psi$ collectively denotes the matter fields.\par Now, the corresponding field equations in the metric approach (varying the action with respect to $g^{\mu\nu}$) yields \cite{n18},\cite{r11}-\cite{r12},
\begin{equation}\label{eq3}
F(R)R_{\mu\nu}-\frac{1}{2}f(R)g_{\mu\nu}-\nabla_\mu\nabla_\nu F(R)+g_{\mu\nu}\Box F(R)=T^m_{\mu\nu}
\end{equation}
where $F=\frac{df}{dR}$ , $\Box =\nabla_\mu\nabla^\mu$, $R_{\mu\nu}$ is the Ricci tensor,
and $T^m_{\mu\nu}$ is the energy momentum tensor. After the contraction of equation (\ref{eq3}), we obtain the following relation
\begin{equation} \label{eq4}
FR-2f+3\Box F=T.
\end{equation}
Using equation (\ref{eq4}) in (\ref{eq1}), we get the standard form of Einstein field equations as
\begin{equation}\label{eq5}
G_{\mu\nu}=R_{\mu\nu}-\frac{1}{2}Rg_{\mu\nu}=T_{\mu\nu}^{\text{eff}}=T_{\mu\nu}^{(g)}+T_{\mu\nu}^m/F,
\end{equation}
where $G_{\mu\nu}$ is the Einstein tensor, $T_{\mu\nu}^{\text{eff}}=T_{\mu\nu}^{(g)}+T_{\mu\nu}^m/F$ is the effective energy-momentum tensor and $T_{\mu\nu}^{(g)}$, due to gravity is given by
\begin{equation}\label{eq6}
T_{\mu\nu}^{(g)}=\frac{1}{F}\left[\nabla_\mu\nabla_\nu F-\frac{1}{4}g_{\mu\nu}(RF+\Box F+T)\right].
\end{equation} 
To obtain some particular wormhole solution, we assume that the stress-energy momentum tensor that threads the wormhole is given by an anisotropic matter distribution as \cite{r13}
$$T_{\mu\nu}=(\rho+p_t)u_\mu u_\nu+p_tg_{\mu\nu}+(p_r-p_t)v_\mu v_\nu,$$ where $v^\mu$ is the unit spacelike vector in the radial direction, $u^\mu$ is the four velocity vector, $\rho(r)$ is the energy density, $p_r(r)$ is the radial pressure measured in the direction of $v^\mu$, and $p_t(r)$ is the transverse pressure measured in the orthogonal direction to $v^\mu$. We shall take the redshift function to be constant which helps to investigate interesting wormhole solutions.\par
Thus, for the wormhole metric (\ref{eq1}), the explicit form of effective field equations (\ref{eq5}) gives \cite{r13}
\begin{equation}\label{eq7}
\frac{\rho}{F}+\frac{J}{F}=\frac{b^\prime}{r^2},
\end{equation}
\begin{equation}\label{eq8}
\frac{p_r}{F}+\frac{1}{F}\left[\left(1-\frac{b}{r}\right)\left(F^{\prime\prime}-F^\prime.\frac{b^\prime r-b}{2r^2(1-b/r)}\right)-J\right]=-\frac{b}{r^3},
\end{equation}
  \begin{equation}\label{eq9}
  \frac{p_t}{F}+\frac{1}{F}\left[\left(1-\frac{b}{r}\right)\frac{F^\prime}{r}-J\right]=-\frac{b^\prime r-b}{2r^3},
  \end{equation}
where the prime denotes a derivative with respect to the radial co-ordinate `$r$'. The term $J=J(r)$ is defined as 
\begin{equation}\label{eq10}
J(r)=\frac{1}{4}(FR+\Box F+T), 
\end{equation}
for notational simplicity. The curvature scalar, $R$ is given by
\begin{equation}\label{eq11}
R=\frac{2b^\prime}{r^2},
\end{equation}
and $\Box F$ has the following expression: 
\begin{equation}\label{eq12}
\Box F=\left(1-\frac{b}{r}\right)\left[F^{\prime\prime}-\frac{b^\prime r-b}{2r^2(1-b/r)}F^\prime+\frac{2F^\prime}{r}\right].
\end{equation}
Now, the gravitational field equations (\ref{eq7})-(\ref{eq9}) can be reorganized to yield as follows:
\begin{equation}\label{eq13}
\frac{Fb^\prime}{r^2}=\rho,
\end{equation}
\begin{equation}\label{eq14}
-\frac{bF}{r^3}+\frac{F^\prime}{2r^2}(b^\prime r-b)-F^{\prime\prime}\left(1-\frac{b}{r}\right)=p_r,
\end{equation}
\begin{equation}\label{eq15}
-\frac{F^\prime}{r}\left(1-\frac{b}{r}\right)+\frac{F}{2r^3}(b-b^\prime r)=p_t.
\end{equation}
This set of coupled ordinary differential equations (\ref{eq13})--(\ref{eq15}) gives the throat function for a given fluid and for a given $F(r)$. As the above set of differential equations is first order in `$b$' so using the throat condition ({\it i.e. }$b(r_0)=r_0$) it is possible to obtain the throat function uniquely.
\section{Wormhole solutions with anisotropic matter field and energy conditions}\label{seciii}
We consider the barotropic equation of state $p_t=\omega\rho$, related to the tangential pressure and the energy density which provides the following differential equation
\begin{equation}\label{eq16}
F^\prime(1-\frac{b}{r})-\frac{F}{2r^2}\left[b-(1+2\omega)b^\prime r\right]=0.
\end{equation}
Now, one may deduce $F(r)$ by imposing a specific shape function, and inverting equation (\ref{eq11}), {\it i.e.} $R(r)$, to find $r(R)$, the specific form of $f(R)$ may be found from trace equation (\ref{eq4}). In the following section, we consider several shape functions usually applied in the literature.
\par
\subsubsection{Shape function: $b(r)=\frac{r_0^n}{r^{n-1}}$ \text{for some} $n>0$ }
 Let us consider the shape function $b(r)=\frac{r_0^n}{r^{n-1}}$ for some $n>0$ (in \cite{r14} the shape function is considered for $n=2$, here we are considering the general case ).
Putting the value of $b(r)$ in equation (\ref{eq16}) we obtain 
\begin{equation}\label{eq17}
F(r)=C_1\left(1-\frac{r_0^n}{r^n}\right)^{\frac{n+(n-1)2\omega}{2n}},
\end{equation} where $C_1$ is arbitrary constant.
The gravitational field equations (\ref{eq13})-(\ref{eq15}) give
\begin{eqnarray}
\rho&=&C_1(1-n)\frac{r_0^n}{r^{n+2}}\left(1-\frac{r_0^n}{r^n}\right)^{\frac{n+(n-1)2\omega}{2n}},\label{eq18}\\
p_r&=&-\frac{C_1}{2}\frac{1}{r^{2(n+1)}}\left(1-\frac{r_0^n}{r^n}\right)^\frac{-n+(n-1)2\omega}{2n}\times \bigg[r^n\bigg\{2-(n+1)(n+(n-1)2\omega)\bigg\}\nonumber\\
&~&+~r_0^n\bigg\{(n+(n-1)2\omega)(n+1+(n-1)\omega)-2\bigg\}\bigg],\label{eq19}\\
p_t&=& C_1\omega(1-n)\frac{r_0^n}{r^{n+2}}\left(1-\frac{r_0^n}{r^n}\right)^{\frac{n+(n-1)2\omega}{2n}}\label{eq20}.
\end{eqnarray}
Now, we have
\begin{eqnarray}
T&=&-\rho+p_r+2p_t\\
&=&C_1\frac{1}{2}\frac{r_0^n}{r^{2(n+1)}}\left(1-\frac{r_0^n}{r^n}\right)^\frac{-n+(n-1)2\omega}{2n}\times\bigg[r^n\bigg\{n^2+2n^2\omega+3n+2\omega-4\omega n-4\bigg\}\nonumber\\\label{eq22}
&~&+~r_0^n\biggl\{4\omega^2n-2n^2\omega^2-3\omega n^2-2\omega^2-n^2-2\omega-3n+5\omega n+4\biggr\}\bigg],
\end{eqnarray}
and using equation(\ref{eq12})
\begin{equation}
\Box F=C_1\frac{n+(n-1)2\omega}{2}\frac{r_0^n}{r^{2(n+1)}}\bigg[r_0^n\biggl\{n-1+(n-1)\omega\biggr\}+r^n(1-n)\bigg].
\end{equation}
The Ricci scalar is given by $R=2(1-n)\frac{r_0^n}{r^{n+2}}$ (using equation (\ref{eq11})) {\it i.e.} $r=\left(\frac{2(1-n)r_0^n}{R}\right)^{\frac{1}{n+2}}$, and hence at the throat we have $$r_0=\left(\frac{2(1-n)}{R_0}\right)^\frac{1}{2}.$$
\begin{figure}
	\includegraphics[height=.35\textheight, width=0.5\textheight]{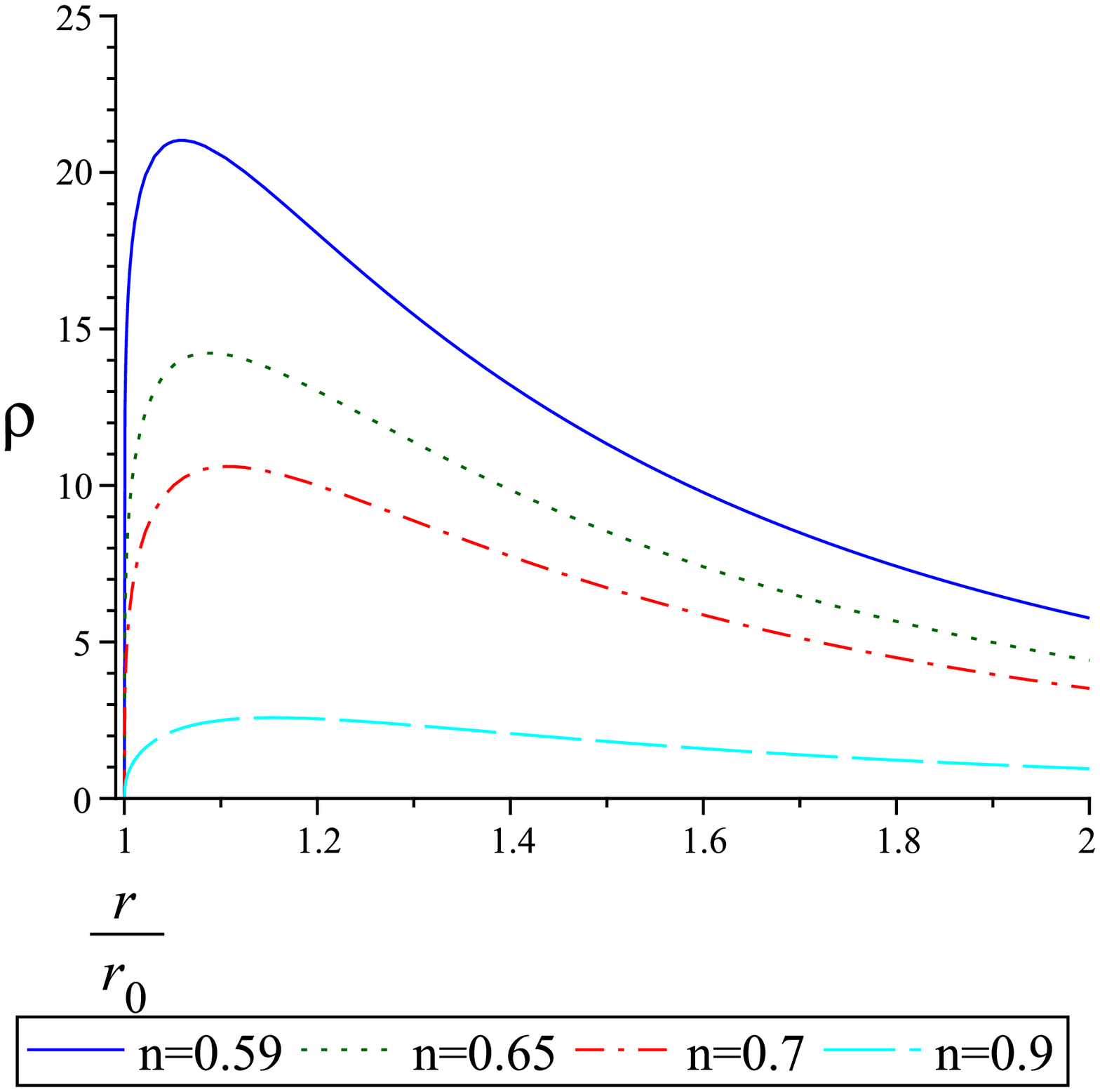}
	\caption{The above figure shows the variations of $\rho$ versus $r/r_0$ for $r_0=0.1$, $\omega=\frac{1}{2}$, $C_1=1$ corresponding to the shape function (1). }
\end{figure}
 \begin{figure}
	\includegraphics[height=0.4\textheight, width=0.5\textheight]{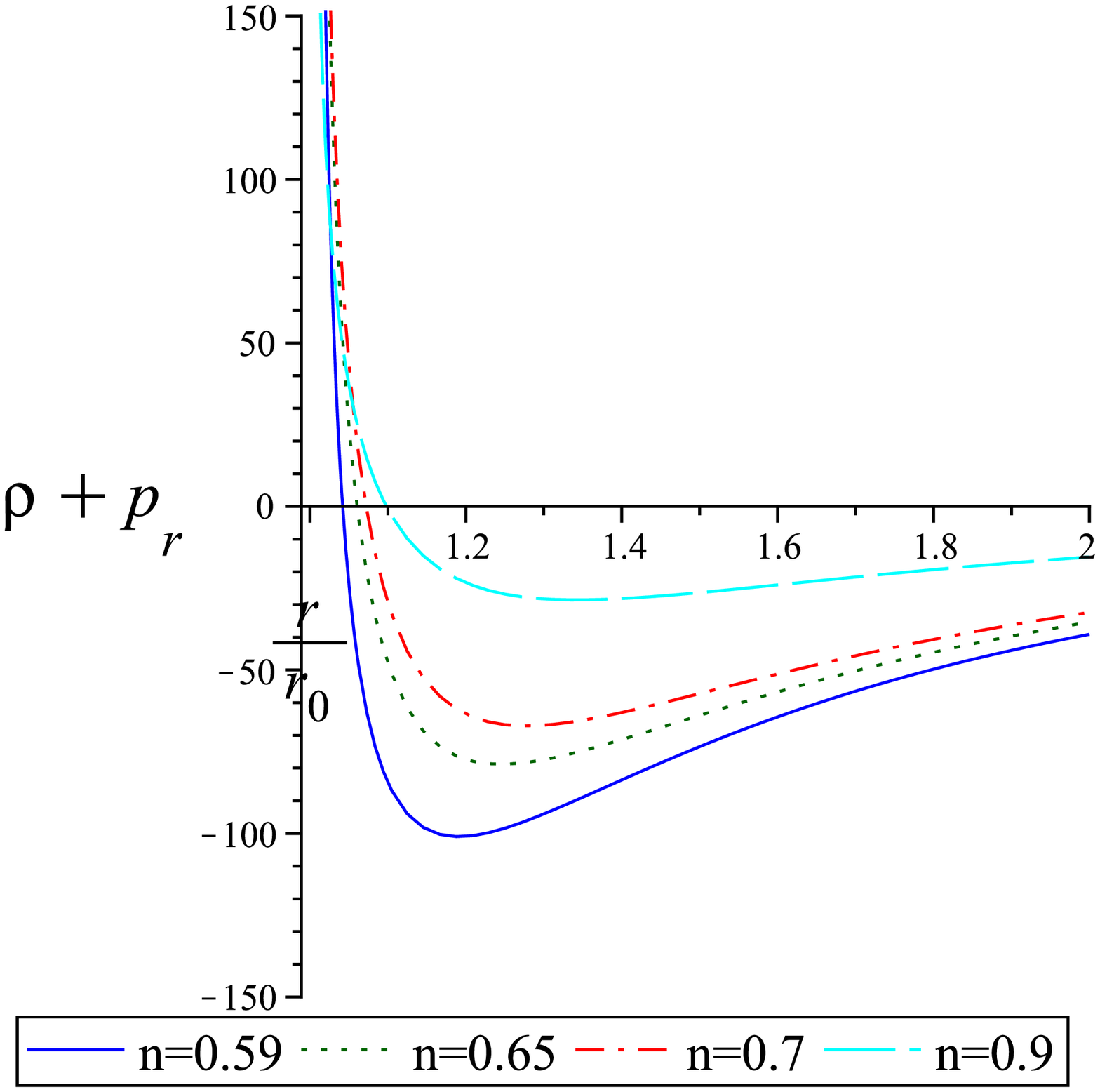}
	\caption{The above figure shows the variations of $\rho+p_r$ versus $r/r_0$ for $r_0=0.1$, $\omega=\frac{1}{2}$, $C_1=1$ corresponding to the shape function (1). }
\end{figure}

After substituting these relations into the equation (\ref{eq4}), the specific form of $f(R)$ is finally given by
 \begin{eqnarray}
f(R)&=&\frac{1}{2}C_1R\bigg\{1-\left(\frac{R}{R_0}\right)^{\frac{n}{n+2}}\bigg\}^{\frac{-n+(n-1)2\omega}{2n}}\Bigg[\left(\frac{R}{R_0}\right)^{\frac{n}{n+2}}\biggl\{-\frac{3}{2}\frac{n+(n-1)2\omega}{2}(1+\omega)-1\nonumber\\
&~&-~\frac{1}{4(1-n)}\left(4-2\omega-3n+5\omega n-2\omega^2-n^2-3\omega n^2+4\omega^2 n-2\omega^2n^2\right)\biggr\}\nonumber\\
&~&+~\biggl\{\frac{3}{2}\frac{n+(n-1)2\omega}{2}+1-\frac{1}{4(1-n)}\left(2\omega n^2+n^2+2\omega+3n-4\omega n-4\right)\biggr\}\Bigg],
\end{eqnarray} as in \cite{r13} if we consider $n=2$.
\par
\par The energy conditions are investigated in terms of principal pressures which are as follows\cite{r16,r18}:\\
$\diamond$ Null energy condition (NEC): $\rho+p_r\geq0$, $\rho+p_t\geq0$,\newline
$\diamond$ Weak energy condition (WEC): $\rho\geq0$, $\rho+p_r\geq0$, $\rho+p_t\geq0$,\newline
$\diamond$ Strong energy condition (SEC): $\rho+p_r\geq0$, $\rho+p_t\geq0$, $\rho+p_r+2p_t\geq0$,\newline
$\diamond$ Dominant energy condition (DEC): $\rho\geq0$, $\rho-|p_r|\geq0$, $\rho-|p_t|\geq0$.\newline

\begin{figure}
	\includegraphics[height=0.4\textheight, width=0.5\textheight]{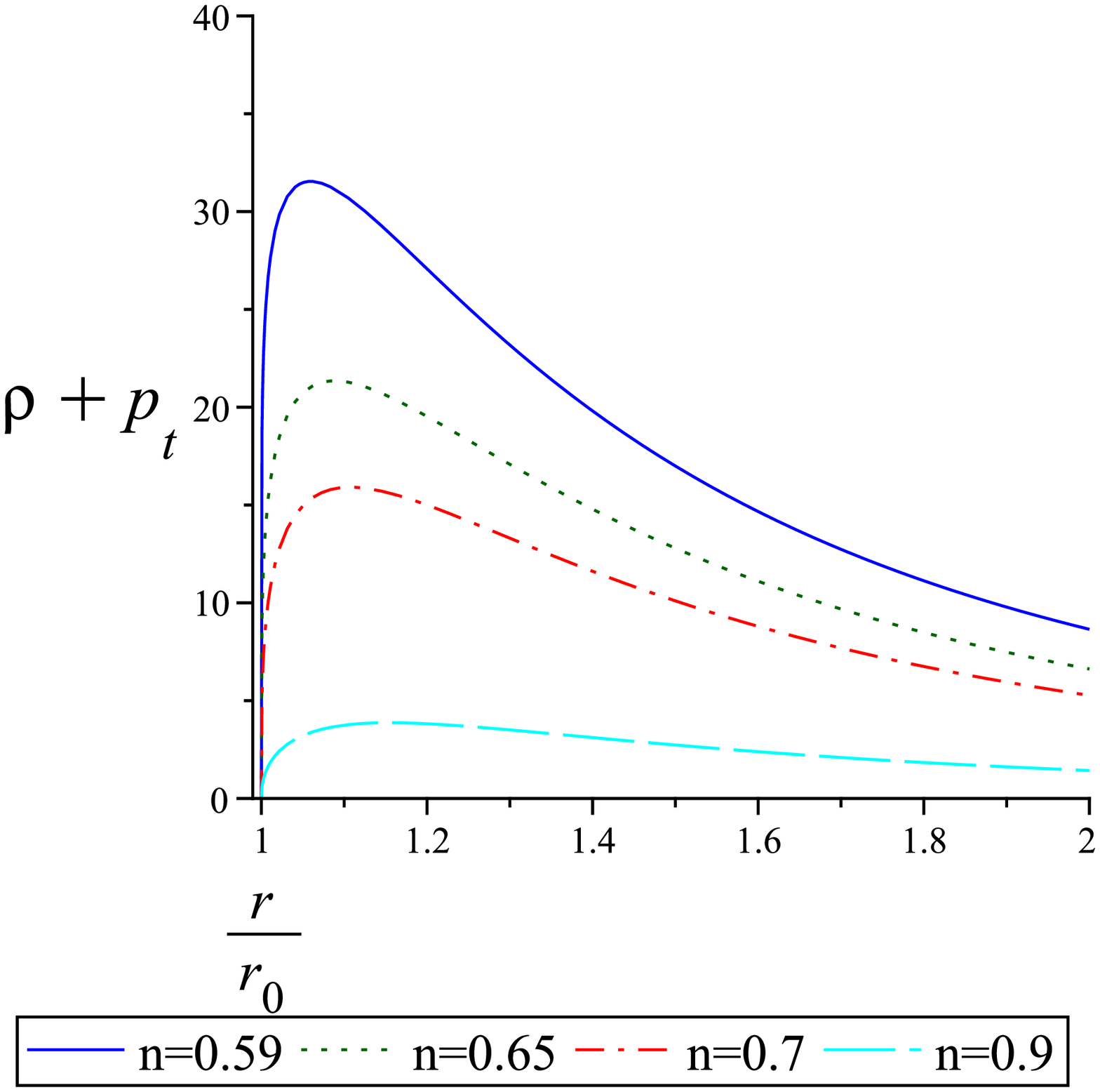}
	\caption{The above figure shows the variations of $\rho+p_t$ versus $r/r_0$ for $r_0=0.1$, $\omega=\frac{1}{2}$, $C_1=1$ corresponding to the shape function (1). }
\end{figure}
\begin{figure}
	\includegraphics[height=0.4\textheight, width=0.5\textheight]{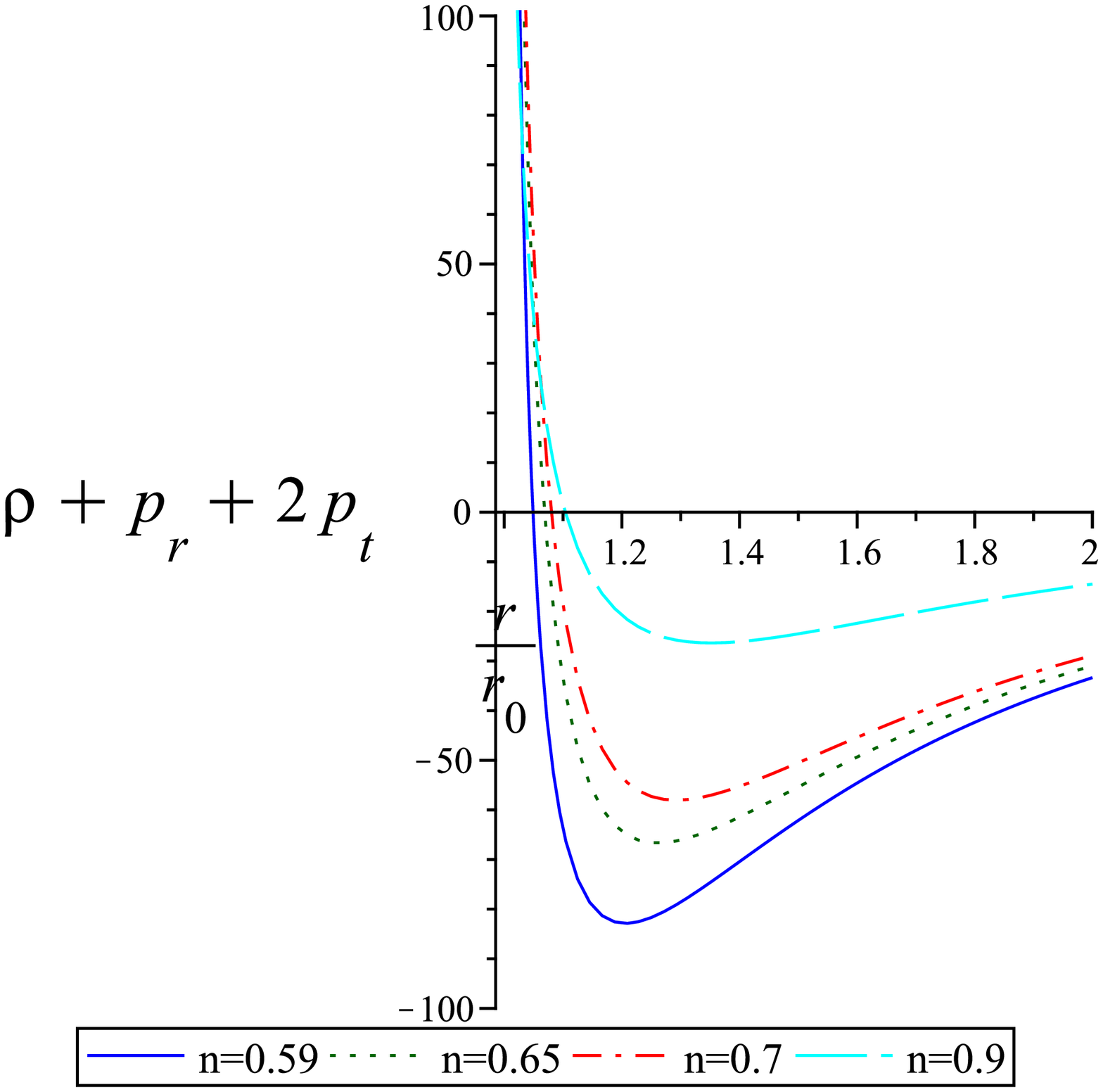}
	\caption{The above figure shows the variations of $\rho+p_r+2p_t$ versus $r/r_0$ for $r_0=0.1$, $\omega=\frac{1}{2}$, $C_1=1$ corresponding the shape function (1). }
\end{figure}
\begin{figure}
	\includegraphics[height=0.4\textheight, width=0.5\textheight]{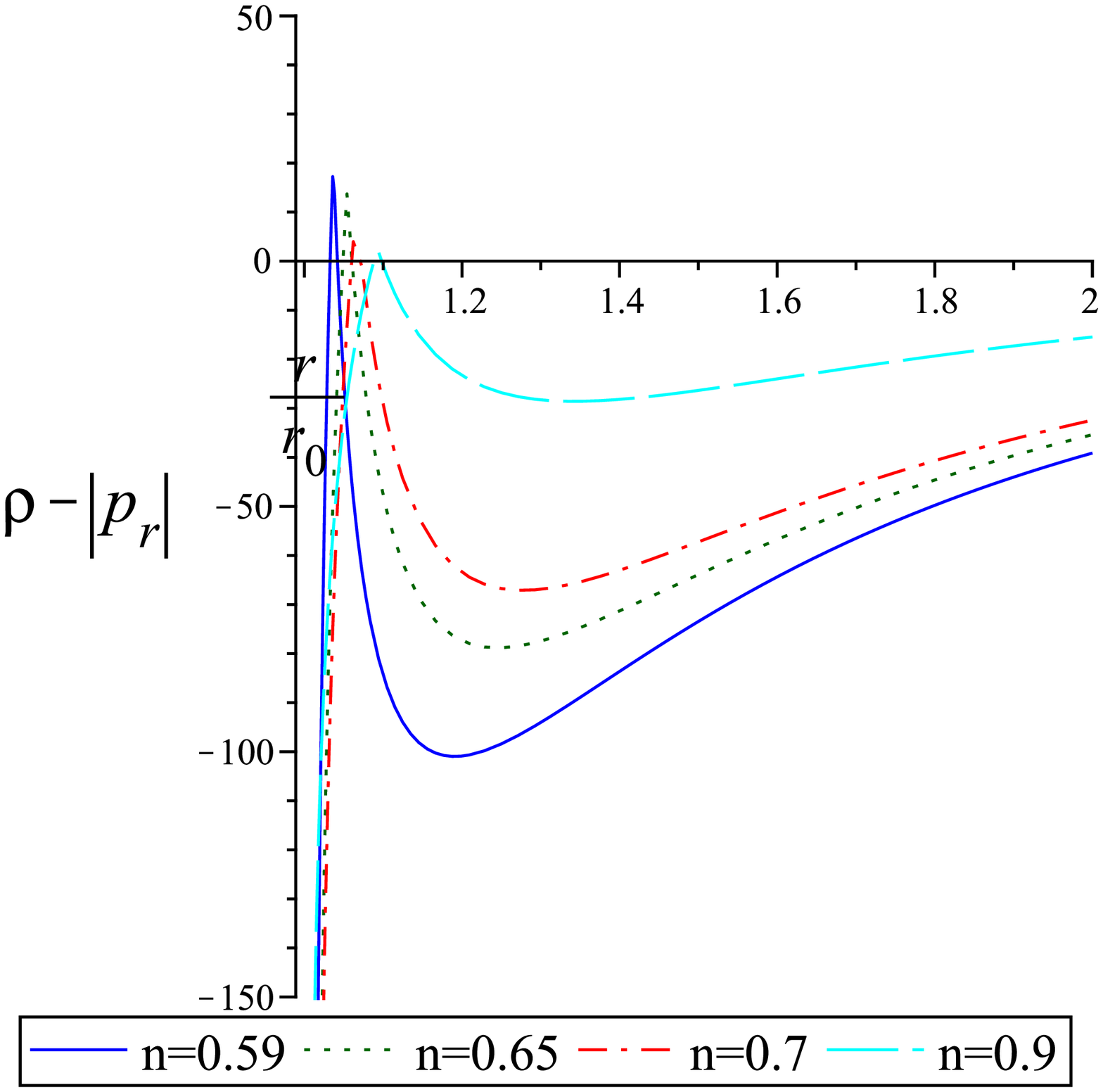}
	\caption{The above figure shows the variations of $\rho-|p_r|$ versus $r/r_0$ for $r_0=0.1$, $\omega=\frac{1}{2}$, $C_1=1$, corresponding the shape function (1). }
\end{figure}
\begin{figure}
	\includegraphics[height=.34\textheight, width=0.5\textheight]{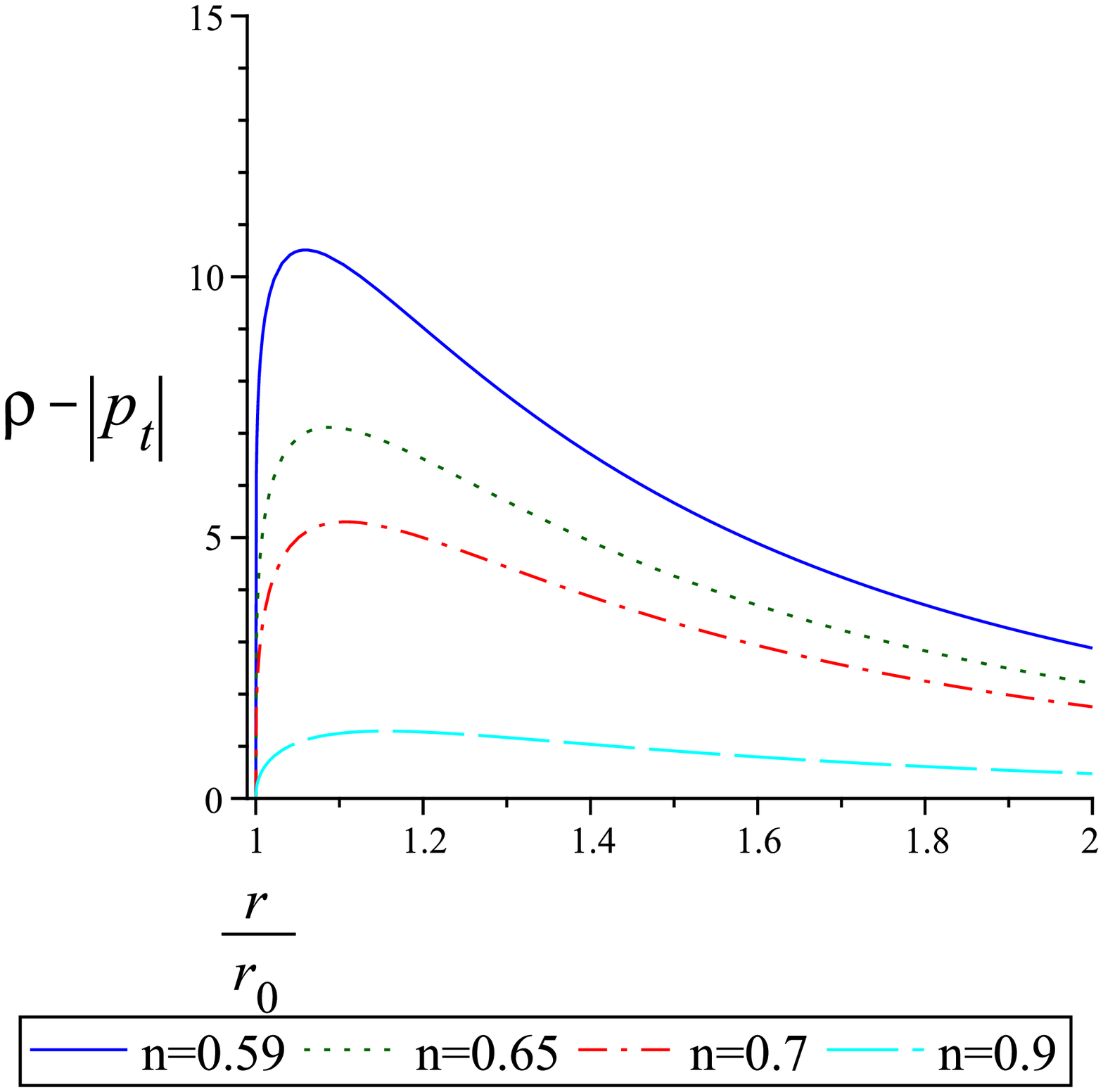}
	\caption{The above figure shows the variations of $\rho-|p_t|$ versus $r/r_0$ for $r_0=0.1$, $\omega=\frac{1}{2}$, $C_1=1$, corresponding the shape function (1). }
\end{figure} 

\par
Considering  $r_0=0.1$, $\omega=\frac{1}{2}$, $C_1=1$ we have drawn the graphs $\rho$ (FIG.1), $\rho+p_r$ (FIG.2), $\rho+p_t$ (FIG.3), $\rho+p_r+2p_t$ (FIG.4), $\rho-|p_r|$ (FIG.5), $\rho-|p_t|$ (FIG.6) with respect to $r/r_0$ for the above shape function.
\subsubsection{Shape function: $b(r)=\frac{r}{1+r-r_0}, 0<r_0<1 $} Let us consider another shape function $b(r)=\frac{r}{1+r-r_0}$, $0<r_0<1$ \cite{r14,r15}.
Putting the value of $b(r)$ in equation (\ref{eq16}) we obtain 
\begin{equation}\label{eq25}
F(r)=C_1r^{\frac{\omega}{r_0}}(r-r_0)^{-\frac{w}{r_0}+\omega+\frac{1}{2}}(1+r-r_0)^{\omega-\frac{1}{2}},
\end{equation}
where $C_1$ is arbitrary constant.
It is useful to write the equation (\ref{eq25}) in the form of 
\begin{equation}
F(r)=C_1X^uY^vZ^s
\end{equation} where $C_1$ is arbitrary constant, and $X$, $Y$, $Z$, $u$, $v$, $s$ are defined as $X=r$, $Y=(r-r_0)$, $Z=(1+r-r_0)$, $u=\frac{\omega}{r_0}$, $v=-\frac{w}{r_0}+\omega+\frac{1}{2}$, $s=\omega-\frac{1}{2}$.\\
Then the gravitational field equations (\ref{eq13})-(\ref{eq15}) become
\begin{eqnarray}
\rho&=&C_1(1-r_0)X^{u-2}Y^vZ^{s-2},
\label{eq31}\\
p_r&=&-C_1\Bigg[X^{u-2}Y^vZ^{s-1}+\frac{u}{2}X^{u-1}Y^{v}Z^{s-2}+\frac{v}{2}X^{u}Y^{v-1}Z^{s-2}+\frac{s}{2}X^{u}Y^{v}Z^{s-3}\nonumber\\
&~&+~u(u-1)X^{u-2}Y^{v+1}Z^{s-1}+v(v-1)X^{u}Y^{v-1}Z^{s-1}+s(s-1)X^{u}Y^{v+1}Z^{s-3}\nonumber\\
&~&+~2uvX^{u-1}Y^{v}Z^{s-1}+2usX^{u-1}Y^{v+1}Z^{s-2}+2vsX^{u}Y^{v}Z^{s-2}\Bigg],\label{eq32}\\
p_t&=& \omega C_1(1-r_0)X^{u-2}Y^vZ^{s-2}.\label{eq33}
\end{eqnarray}
Now, we have
\begin{eqnarray}
T&=&-\rho+p_r+2p_t\nonumber\\
&=&C_1\Bigg[(2\omega-1)(1-r_0)X^{u-2}Y^vZ^{s-2}-X^{u-2}Y^vZ^{s-1}-\frac{u}{2}X^{u-1}Y^{v}Z^{s-2}-\frac{v}{2}X^{u}Y^{v-1}Z^{s-2}\nonumber\\
&~&-~\frac{s}{2}X^{u}Y^{v}Z^{s-3}-u(u-1)X^{u-2}Y^{v+1}Z^{s-1}-v(v-1)X^{u}Y^{v-1}Z^{s-1}-s(s-1)X^{u}Y^{v+1}Z^{s-3}\nonumber\\
&~&-~2uvX^{u-1}Y^{v}Z^{s-1}-2usX^{u-1}Y^{v+1}Z^{s-2}-2vsX^{u}Y^{v}Z^{s-2}\Bigg],
\end{eqnarray}
\begin{figure}
	
	\includegraphics[ width=.4\textheight]{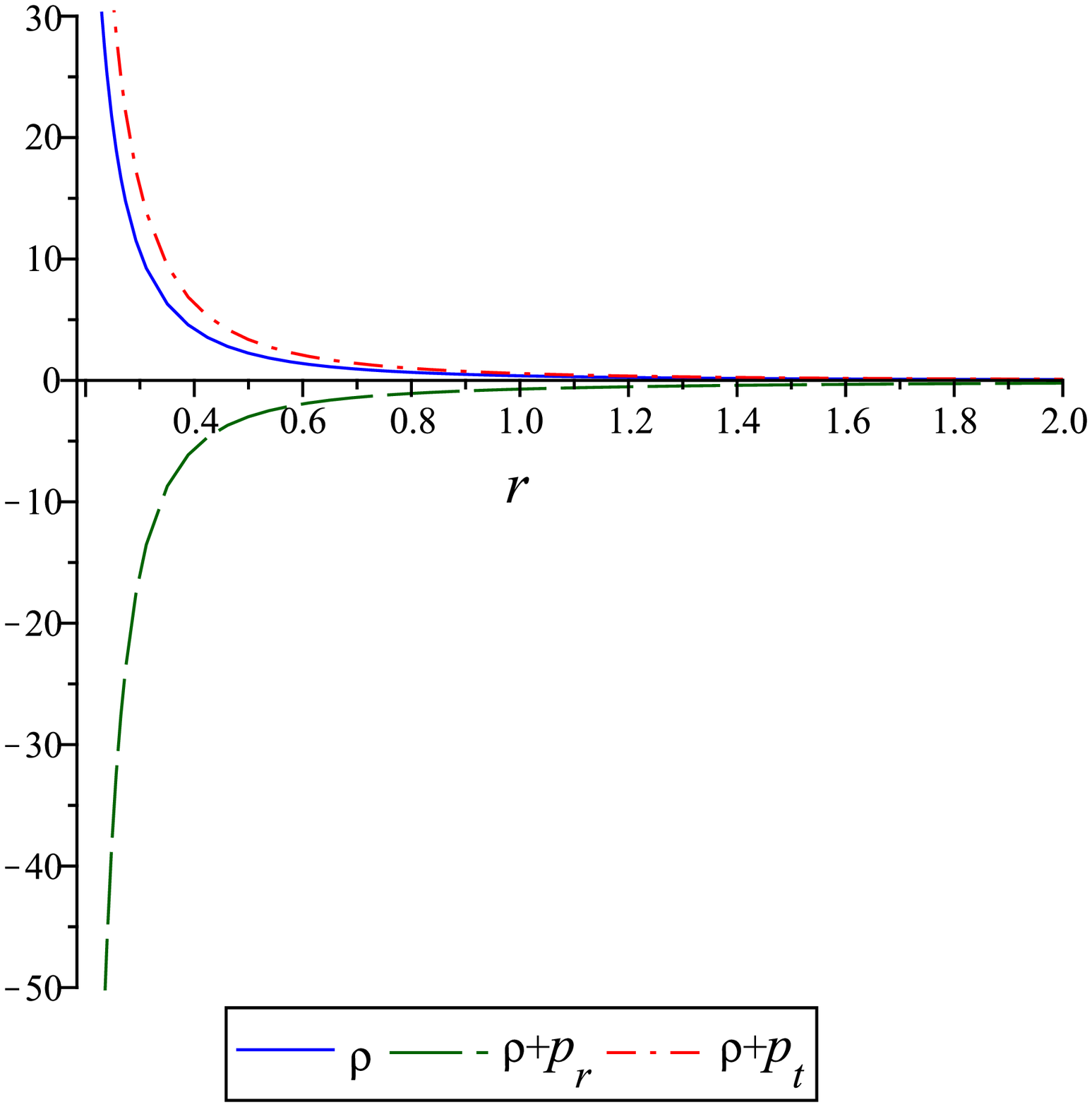}
	\caption{ Variation of $\rho$ (solid curve), $\rho+p_r$ (longdashed curve) and $\rho+p_t$ (dashdot curve) for $C_1=1$, $r_0=0.1$, $\omega=1/2$  corresponding to the shape function (2).}
\end{figure}
\begin{figure}
	
	\includegraphics[ width=.4\textheight]{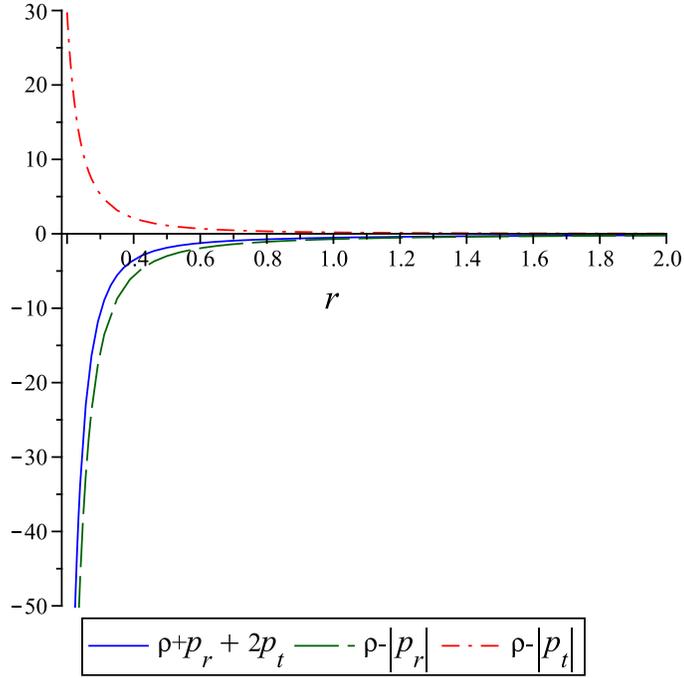}
	\caption{ Variation of $\rho+p_r+2p_t$ (solid curve), $\rho-|p_r|$ (longdashed curve) and $\rho-|p_t|$ (dashdot curve) for $C_1=1$, $r_0=0.1$, $\omega=1/2$  corresponding to the shape function (2).}
	
\end{figure}
and using equation(\ref{eq12})
\begin{eqnarray}
\Box F&=&C_1\Bigg[u(u-1)X^{u-2}Y^{v+1}Z^{s-1}+v(v-1)X^{u}Y^{v-1}Z^{s-1}+s(s-1)X^{u}Y^{v+1}Z^{s-3}+2uvX^{u-1}Y^{v}Z^{s-2}\nonumber\\
&~&+~2usX^{u-1}Y^{v+1}Z^{s-2}+2vsX^{u}Y^{v}Z^{s-2}+\frac{u}{2}X^{u-1}Y^{v-1}Z^{s-1}+\frac{v}{2}X^{u}Y^{v-2}Z^{s-1}\nonumber\\
&~&+~\frac{s}{2}X^{u}Y^{v-1}Z^{s-2}+2uX^{u-2}Y^{v}Z^{s}+2vX^{u-1}Y^{v-1}Z^{s}+2sX^{u-1}Y^{v}Z^{s-1}\Bigg].
\end{eqnarray}
The Ricci scalar is given by $R=2(1-r_0)X^{-2}Z^{-2}$ (using equation (\ref{eq11})). 
After substituting these relations into the equation (\ref{eq4}), the specific form of $f(r)$ is finally given by
\begin{eqnarray}
f(r)&=&C_1\Bigg[4u(u-1)X^{u-2}Y^{v+1}Z^{s-1}+(8uv+6s)X^{u-1}Y^{v}Z^{s-1}+8usX^{u-1}Y^{v+1}Z^{s-2}+8vsX^{u}Y^{v}Z^{s-2}\nonumber\\
&~&+~3v(v-1)X^{u}Y^{v-1}Z^{s-1}+4s(s-1)X^{u}Y^{v+1}Z^{s-3}+\frac{3u}{2}X^{u-1}Y^{v-1}Z^{s-1}+\frac{3v}{2}X^{u}Y^{v-2}Z^{s-1}\nonumber\\
&~&+\left(\frac{3s}{2}+v(v-1)\right)X^{u}Y^{v-1}Z^{s-2}+6uX^{u-2}Y^{v}Z^{s}+6vX^{u-1}Y^{v-1}Z^{s}\nonumber\\
&~&+(-2\omega+3)(1-r_0)X^{u-2}Y^{v}Z^{s-2}+~X^{u-2}Y^{v}Z^{s-1}+\frac{u}{2}X^{u-1}Y^{v}Z^{s-2}\nonumber\\
&~& +\frac{v}{2}X^{u}Y^{v-1}Z^{s-2}+\frac{s}{2}X^{u}Y^{v}Z^{s-2}\Bigg].
\end{eqnarray}

 Using the same value of the parameters as in above, we have drawn the graphs $\rho$, $\rho+p_r$ and $\rho+p_t$ (FIG.7), $\rho+p_r+2p_t$, $\rho-|p_r|$ and $\rho-|p_t|$ (FIG.8) with respect to $r$ for this shape function.  
\subsubsection{Shape function: $b(r)=re^{-2(r-r_0)}$} Lastly, let us consider another shape function $b(r)=re^{-2(r-r_0)}$ \cite{r14}.
Putting the value of $b(r)$ in equation (\ref{eq16}) we obtained 
\begin{equation}\label{eq31}
F(r)=\left(1-e^{-2(r-r_0)}\right)^{\omega+\frac{1}{2}}e^{\int\frac{-\omega }{r\left(e^{2(r-r_0)}-1\right)}dr}.
\end{equation}
Then the gravitational field equations (\ref{eq13})-(\ref{eq15}) become

\begin{eqnarray}
\rho&=&\frac{1-2r}{r^2}e^{-2(r-r_0)}\left(1-e^{-2(r-r_0)}\right)^{\omega+\frac{1}{2}}e^{\int\frac{-\omega }{r\left(e^{2(r-r_0)}-1\right)}dr},\label{eq33}\\
p_r&=&-\rho\bigg[\frac{1}{1-2r}+\frac{r^2}{(1-2r)}\frac{(1+2\omega)r-\omega}{r\left(e^{2(r-r_0)}-1\right)}+\bigg\{\frac{(1+2\omega)^2-2(1+2\omega)e^{2(r-r_0)}}{\left(e^{2(r-r_0)}-1\right)^2}\nonumber\\
&~&+~\frac{\omega^2+\omega\left((2r+1)e^{2(r-r_0)}-1\right)}{r^2\left(e^{2(r-r_0)}-1\right)^2}-\frac{2(1+2\omega)\omega}{r\left(e^{2(r-r_0)}-1\right)^2}
\bigg\}\frac{r^2e^{2(r-r_0)}}{(1-2r)}\bigg],\label{eq35}\\
p_t&=& \omega \frac{1-2r}{r^2}e^{-2(r-r_0)}\left(1-e^{-2(r-r_0)}\right)^{\omega+\frac{1}{2}}e^{\int\frac{-\omega }{r\left(e^{2(r-r_0)}-1\right)}dr}.
\end{eqnarray}

Now, we have
\begin{eqnarray}
T&=&-\rho+p_r+2p_t\nonumber\\
&=&\bigg[(1-2\omega)\frac{1-2r}{r^2}e^{-2(r-r_0)}+\bigg\{\frac{-1}{r^2e^{2(r-r_0)}}-\frac{1}{e^{2(r-r_0)}}\frac{(1+2\omega)r-\omega}{r\left(e^{2(r-r_0)}-1\right)}+\bigg\{\frac{(1+2\omega)^2-2(1+2\omega)e^{2(r-r_0)}}{\left(e^{2(r-r_0)}-1\right)^2}\nonumber\\
&~&+~ \frac{\omega^2+\omega\left((2r+1)e^{2(r-r_0)}-1\right)}{r^2\left(e^{2(r-r_0)}-1\right)^2}-\frac{2(1+2\omega)\omega}{r\left(e^{2(r-r_0)}-1\right)^2}
\bigg\}\left(1-e^{-2(r-r_0)}\right)\bigg\}\bigg]F(r),
\end{eqnarray}
and using equation(\ref{eq12})
\begin{eqnarray}
\Box F&=&\bigg[\bigg\{\frac{(1+2\omega)^2-2(1+2\omega)e^{2(r-r_0)}}{\left(e^{2(r-r_0)}-1\right)^2}+
\frac{\omega^2+\omega\left((2r+1)e^{2(r-r_0)}-1\right)}{r^2\left(e^{2(r-r_0)}-1\right)^2}-\frac{2(1+2\omega)\omega}{r\left(e^{2(r-r_0)}-1\right)^2}
\bigg\}\nonumber\\
&~&~\left(1-e^{-2(r-r_0)}\right)
+\frac{\left((1+2\omega)r-\omega\right)}{r\left(e^{2(r-r_0)}-1\right)}\frac{\left(-r+2(\left(e^{2(r-r_0)}-1\right))\right)}{re^{2(r-r_0)}}\bigg]F(r).
\end{eqnarray}

Now, Ricci scalar is given by $R=\frac{2(1-2r)e^{2(r-r_0)}}{r^2}$ (using equation(\ref{eq11})). 
After substituting these relations into the relation (\ref{eq4}), the specific form of $f(r)$ is finally given by
\begin{eqnarray}
f(r)&=&\frac{1}{2}\bigg[\frac{2(1-2r)e^{2(r-r_0)}}{r^2}+2\bigg\{\frac{(1+2\omega)^2-2(1+2\omega)e^{2(r-r_0)}}{\left(e^{2(r-r_0)}-1\right)^2}+\frac{\omega^2+\omega\left((2r+1)e^{2(r-r_0)}-1\right)}{r^2\left(e^{2(r-r_0)}-1\right)^2}\nonumber\\
&~&-~\frac{2(1+2\omega)\omega}{r\left(e^{2(r-r_0)}-1\right)^2}
\bigg\}\left(1-e^{-2(r-r_0)}\right)+\frac{\left((1+2\omega)r-\omega\right)}{r\left(e^{2(r-r_0)}-1\right)}\frac{\left(-r+2(\left(e^{2(r-r_0)}-1\right))\right)}{re^{2(r-r_0)}}\nonumber\\
&~&-~(1-2\omega)\frac{1-2r}{r^2}e^{-2(r-r_0)}+\frac{1}{r^2e^{2(r-r_0)}}+\frac{1}{e^{2(r-r_0)}}\frac{(1+2\omega)r-\omega}{r\left(e^{2(r-r_0)}-1\right)}
\bigg]F(r).
\end{eqnarray}
\begin{figure}[h]
	\centering
	\begin{minipage}{.32\textwidth}
		\centering
		\includegraphics[width=.6\linewidth]{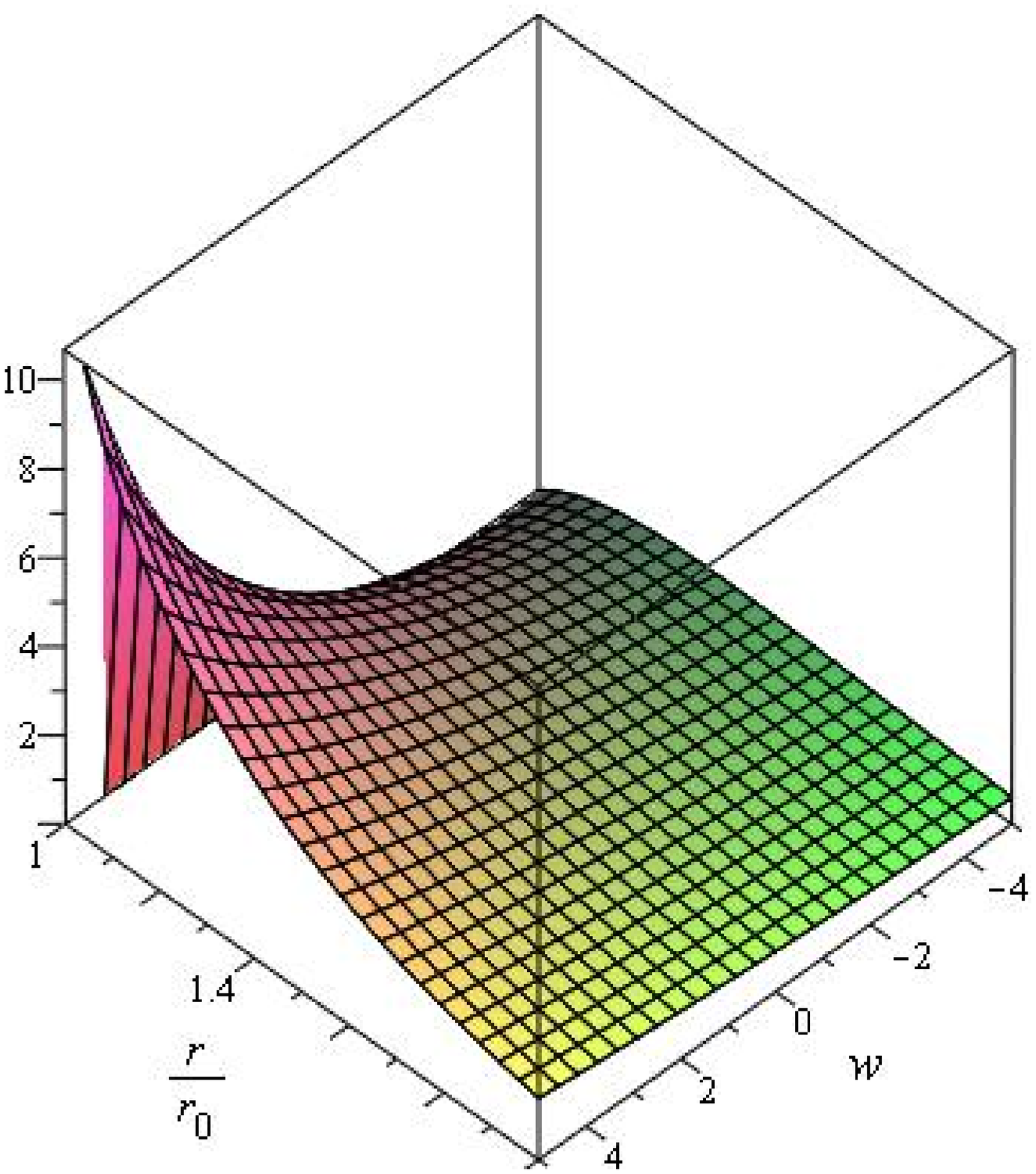}
		\centering FIG.9(A)
	\end{minipage}
	\begin{minipage}{.32\textwidth}
		\centering
		\includegraphics[width=.6\linewidth]{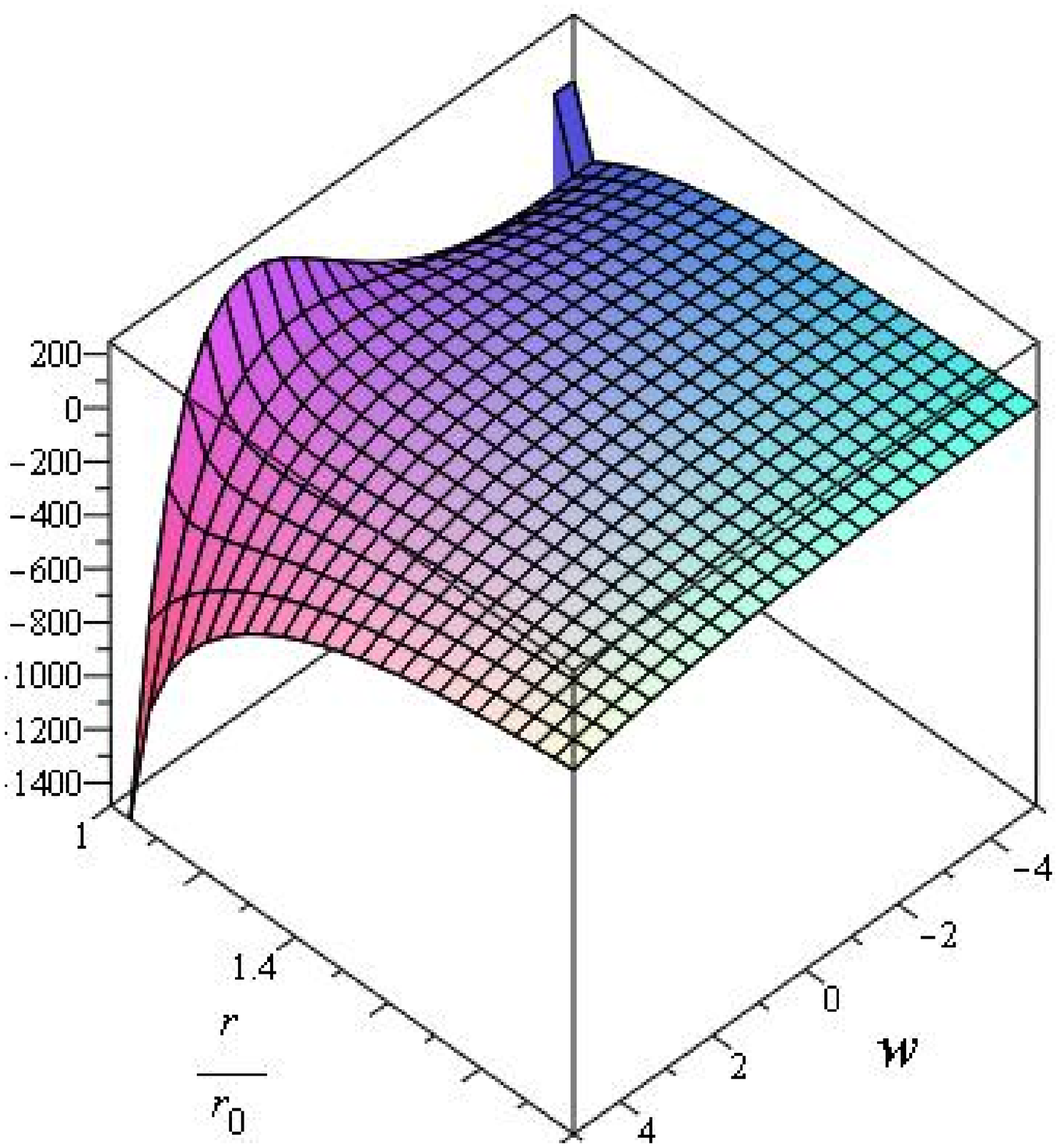}
		\centering FIG.9(B)
	\end{minipage}
	\begin{minipage}{.32\textwidth}
		\centering
		\includegraphics[width=.6\linewidth]{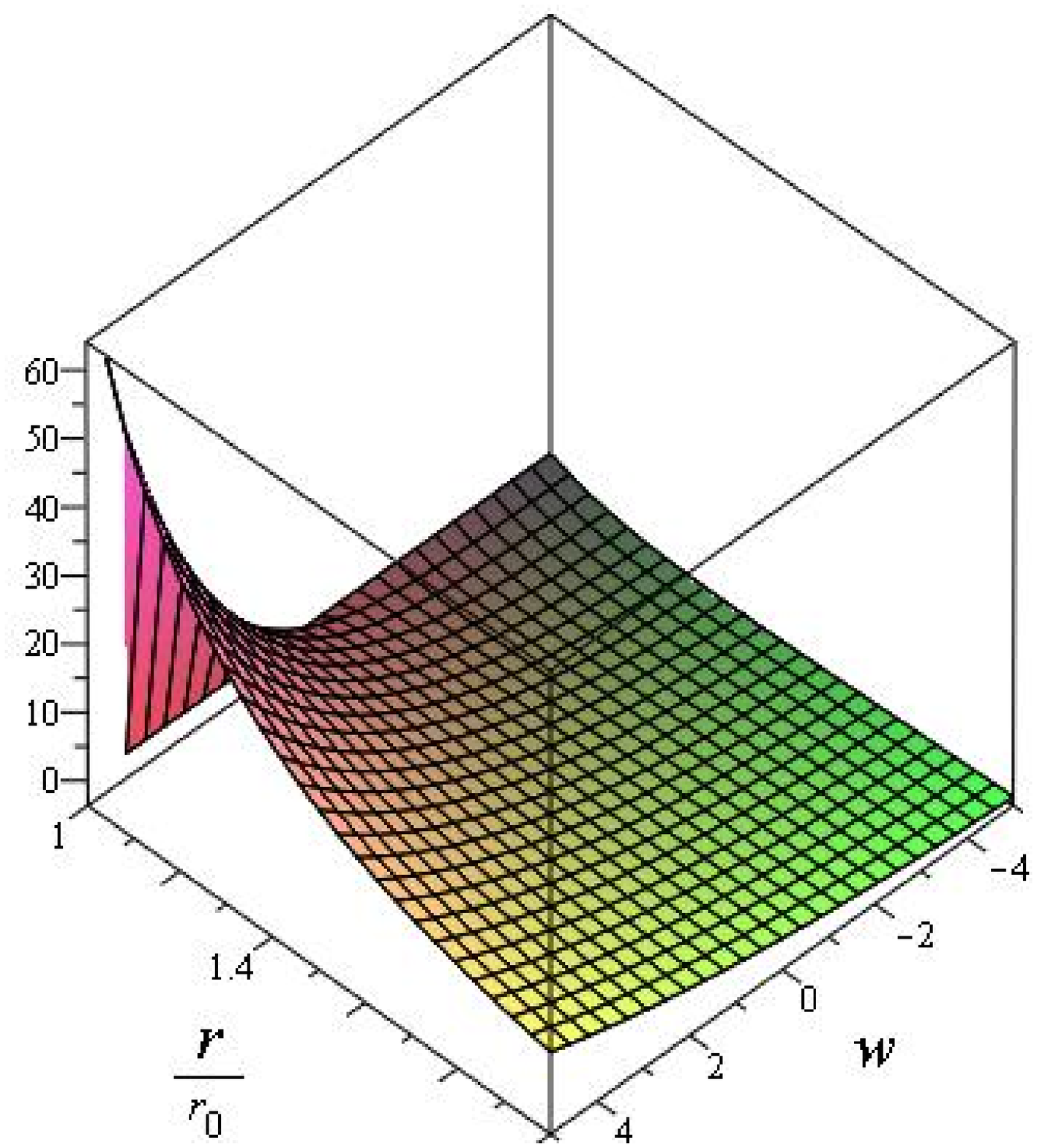}
		\centering FIG.9(C)
	\end{minipage}
\end{figure}
\begin{figure}[h]
	\centering
	\begin{minipage}{.32\textwidth}
		\centering
		\includegraphics[width=.6\linewidth]{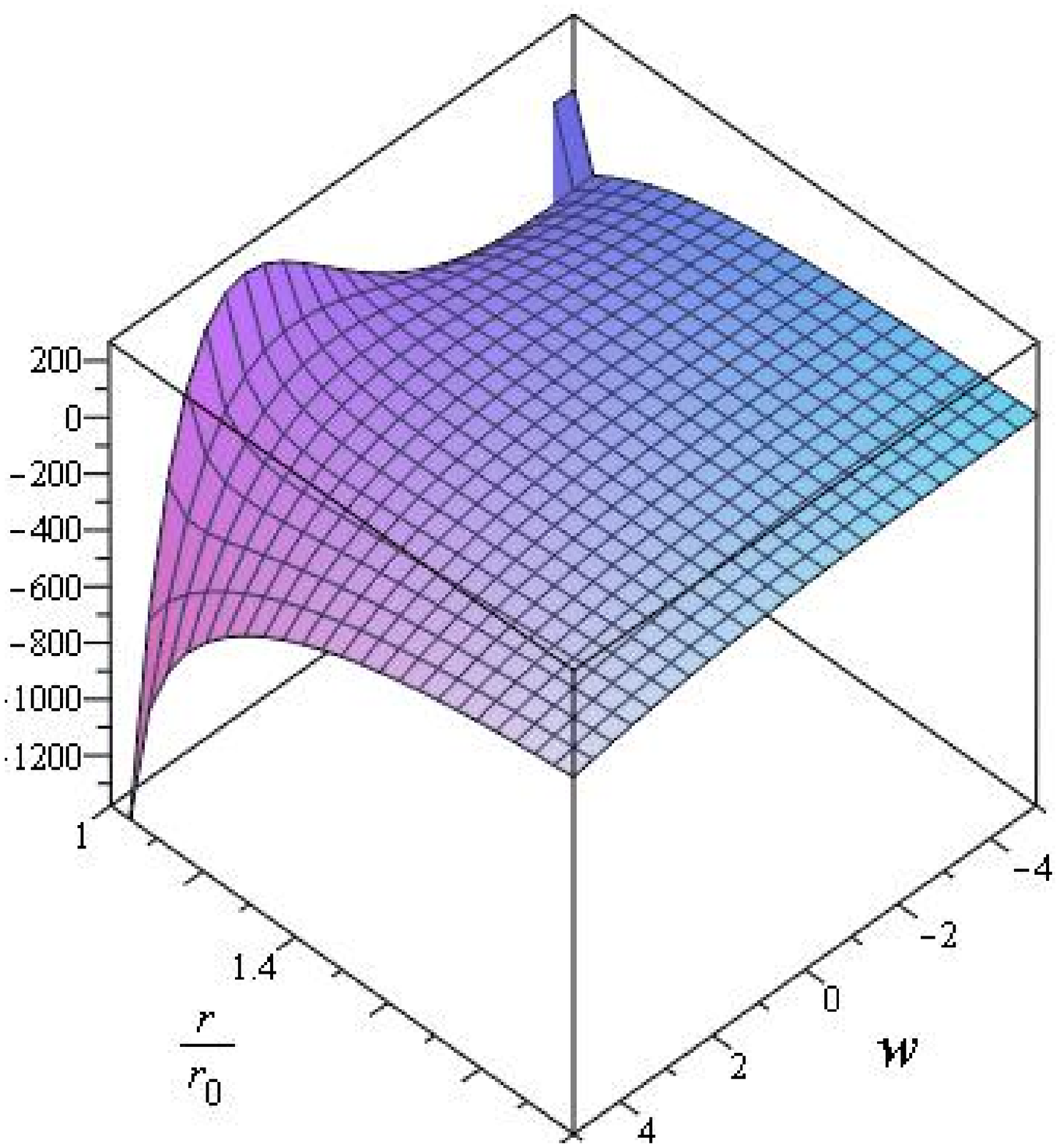}
		\centering FIG.9(D)
	\end{minipage}
	\begin{minipage}{.32\textwidth}
		\centering
		\includegraphics[width=.6\linewidth]{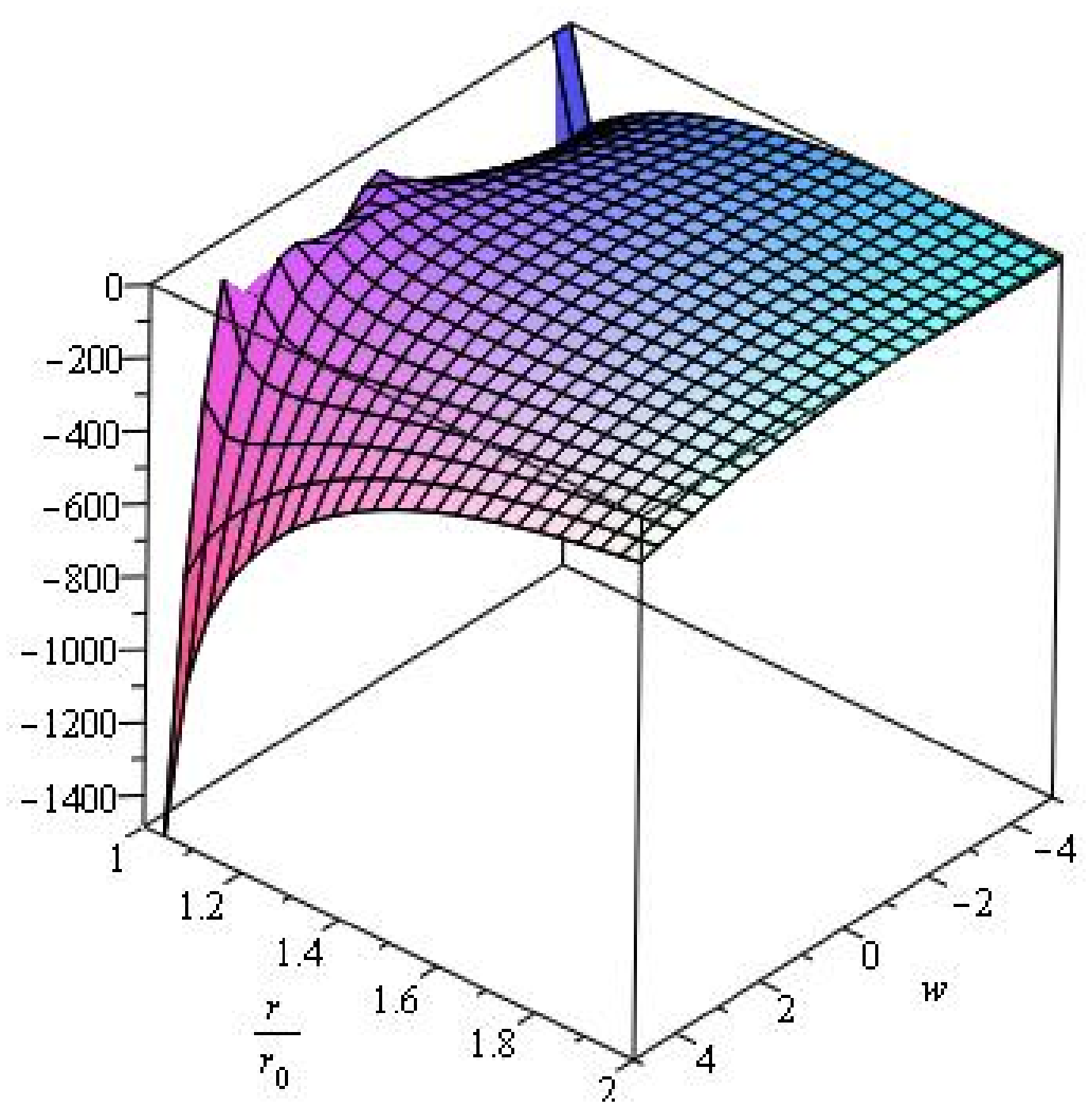}
		\centering FIG.9(E)
	\end{minipage}
	\begin{minipage}{.32\textwidth}
		\centering
		\includegraphics[width=.6\linewidth]{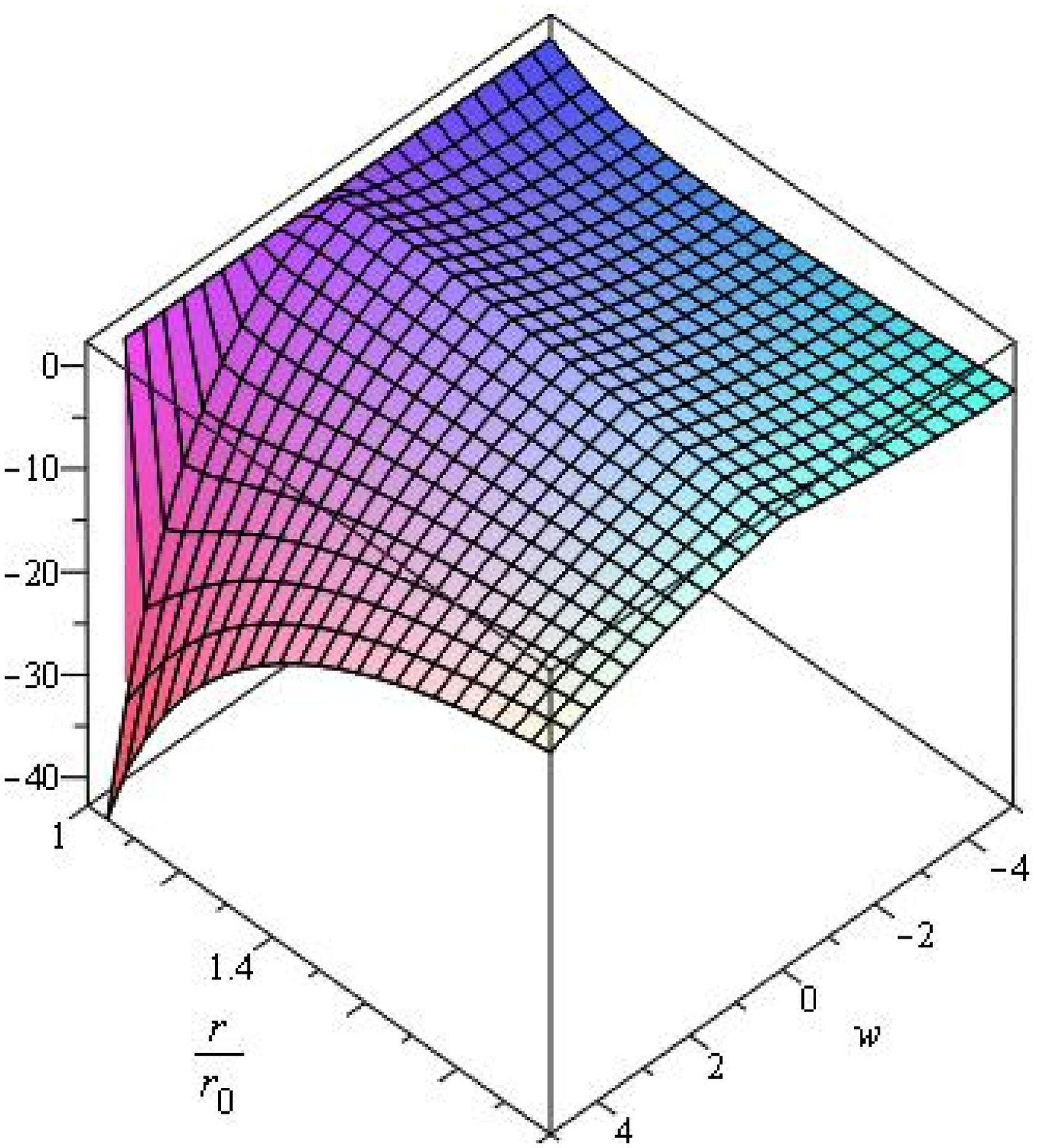}
		\centering FIG.9(F)
	\end{minipage}
	\caption{Behavior of $\rho$ (FIG.9(A)), $\rho+p_r$ (FIG.9(B)), $\rho+p_t$ (FIG.9(C)), $\rho+p_r+2p_t$ (FIG.9(D)), $\rho-|p_r|$ (FIG.9(E)) and $\rho-|p_t|$ (FIG.9(F)) diagrams have been plotted  against $r/r_0$ and $\omega$ for shape function(1) when $C_1=1$, $-5\leq\omega\leq5$, $n=0.9$ and $r_0=0.1$.}	\label{fig9.1}
\end{figure}

\begin{figure}[!htb]
	\centering
	\begin{minipage}{.32\textwidth}
		\centering
		\includegraphics[width=.6\linewidth]{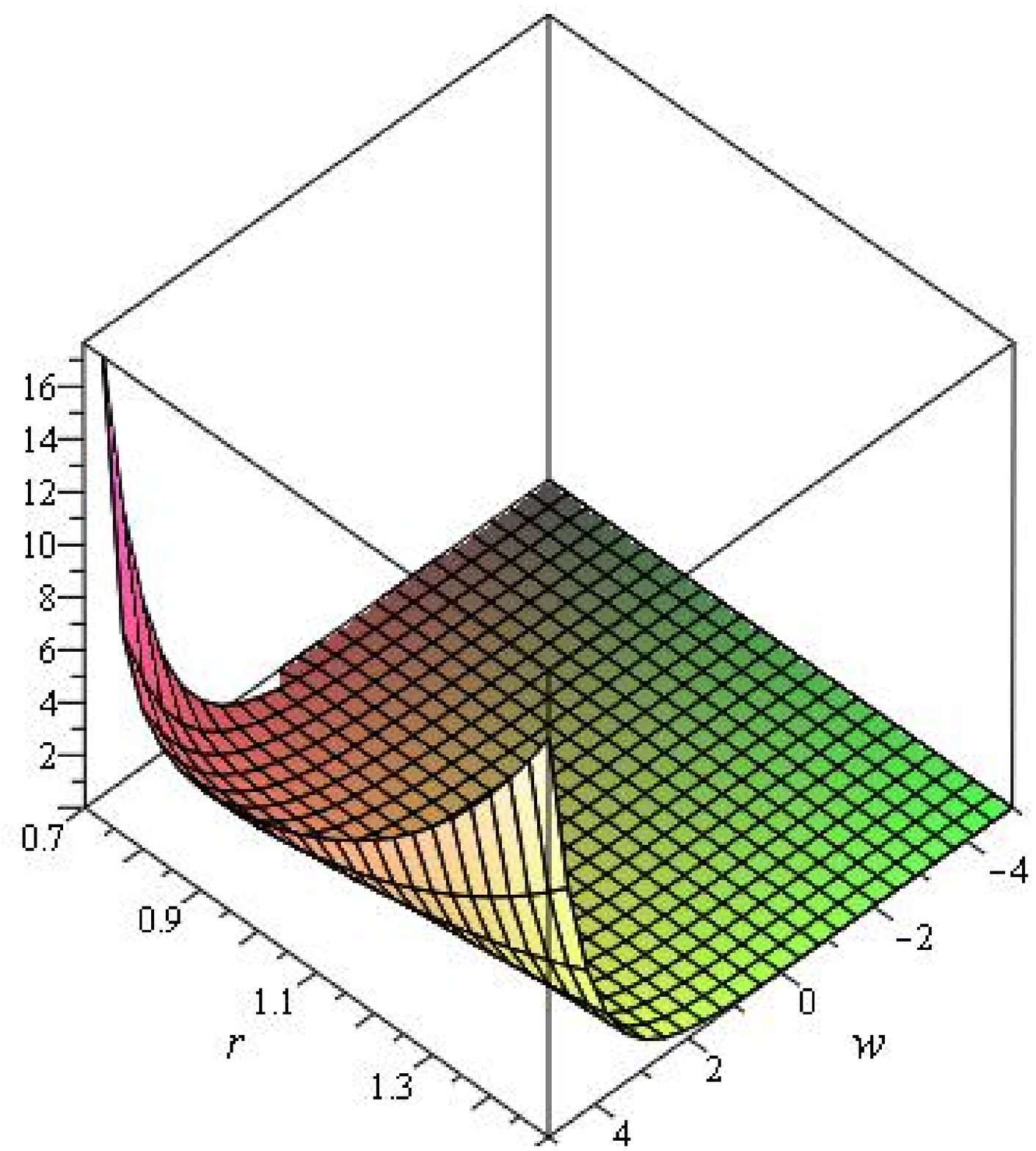}
		\centering FIG.10(A)
	\end{minipage}
	\begin{minipage}{.32\textwidth}
		\centering
		\includegraphics[width=.6\linewidth]{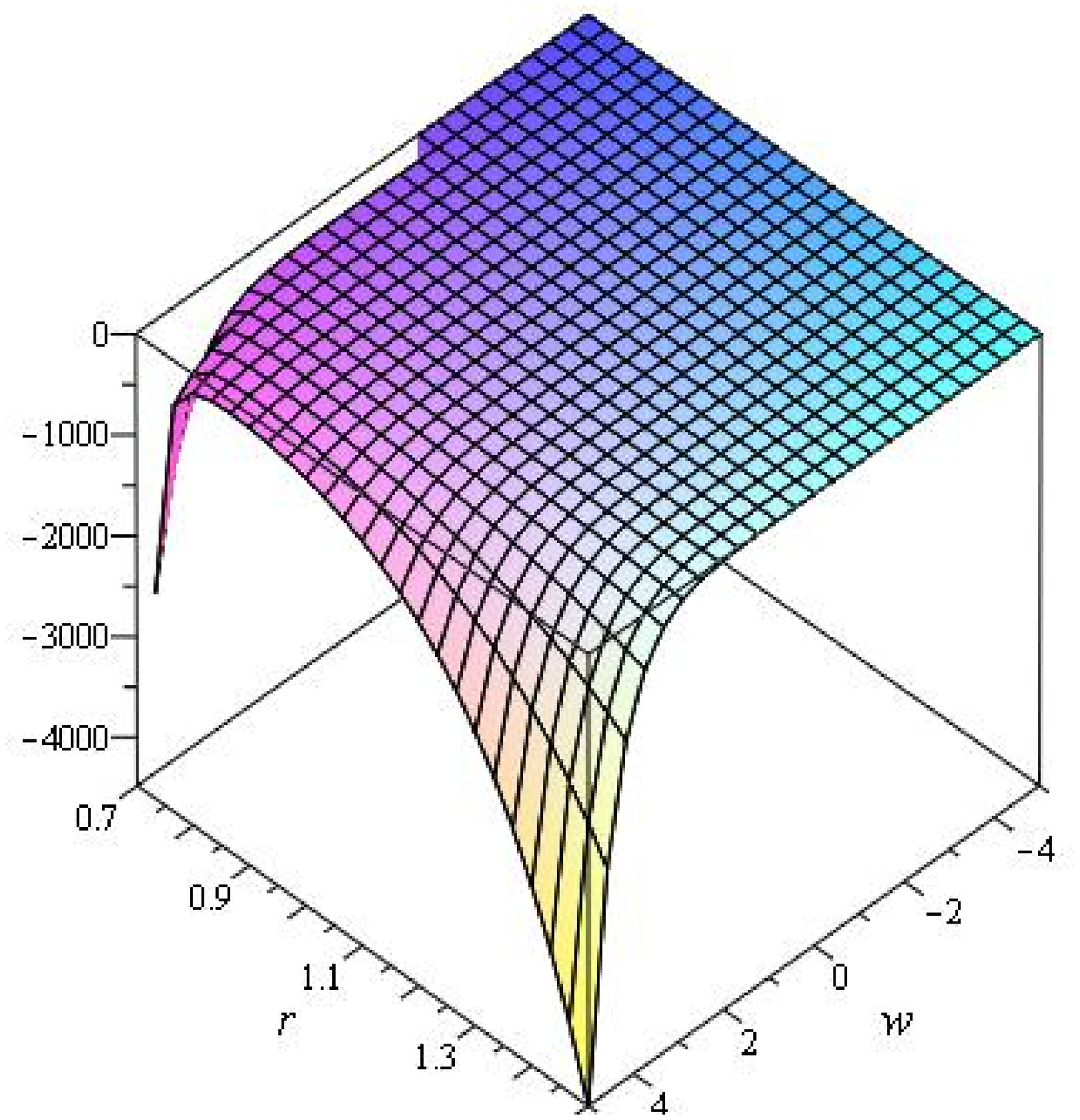}
		\centering FIG.10(B)
	\end{minipage}
	\begin{minipage}{.32\textwidth}
		\centering
		\includegraphics[width=.6\linewidth]{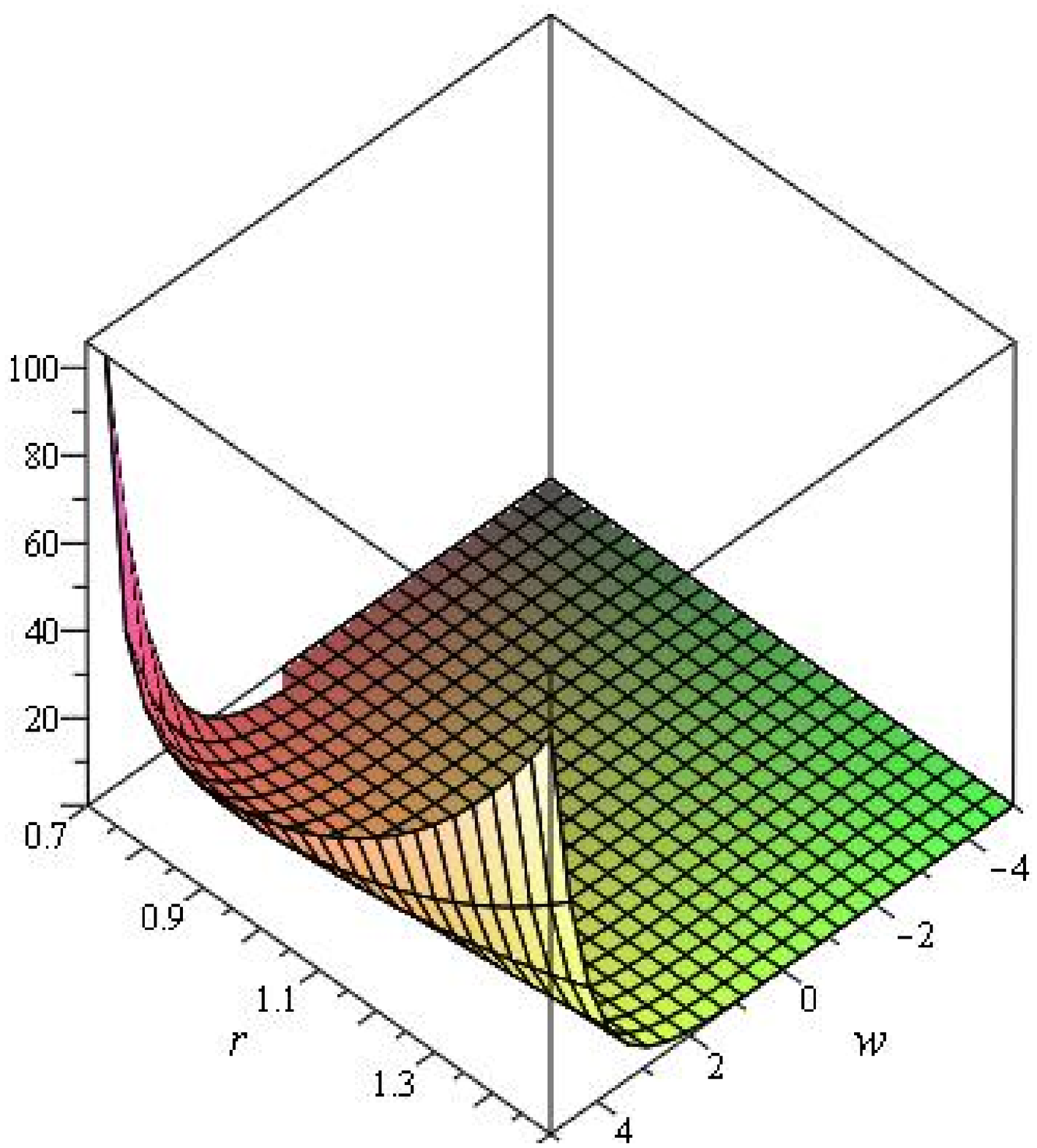}
		\centering FIG.10(C)
	\end{minipage}
\end{figure}
\begin{figure}[!htb]
	\centering
	\begin{minipage}{.32\textwidth}
		\centering
		\includegraphics[width=.6\linewidth]{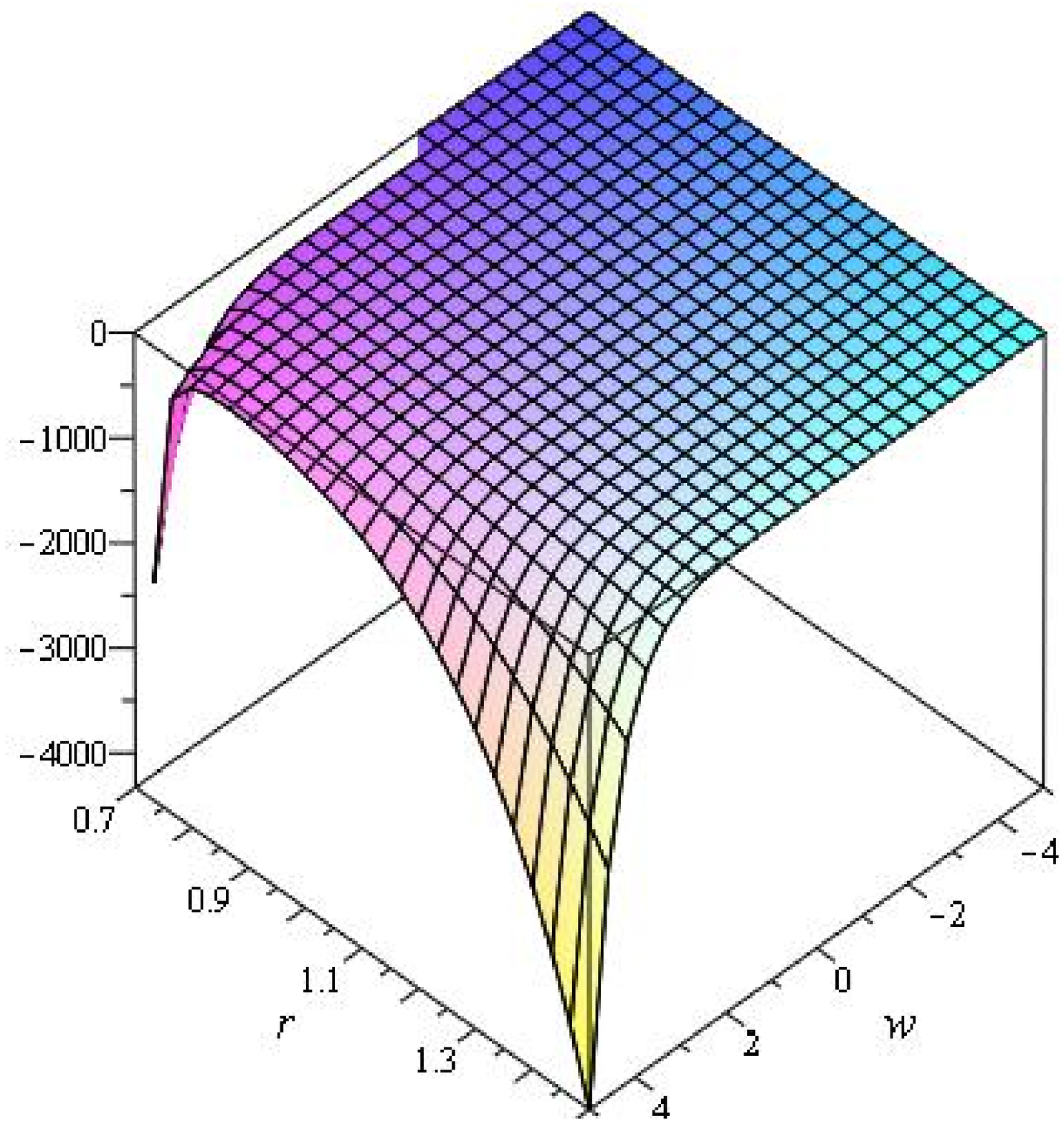}
		\centering FIG.10(D)
	\end{minipage}
	\begin{minipage}{.32\textwidth}
		\centering
		\includegraphics[width=.6\linewidth]{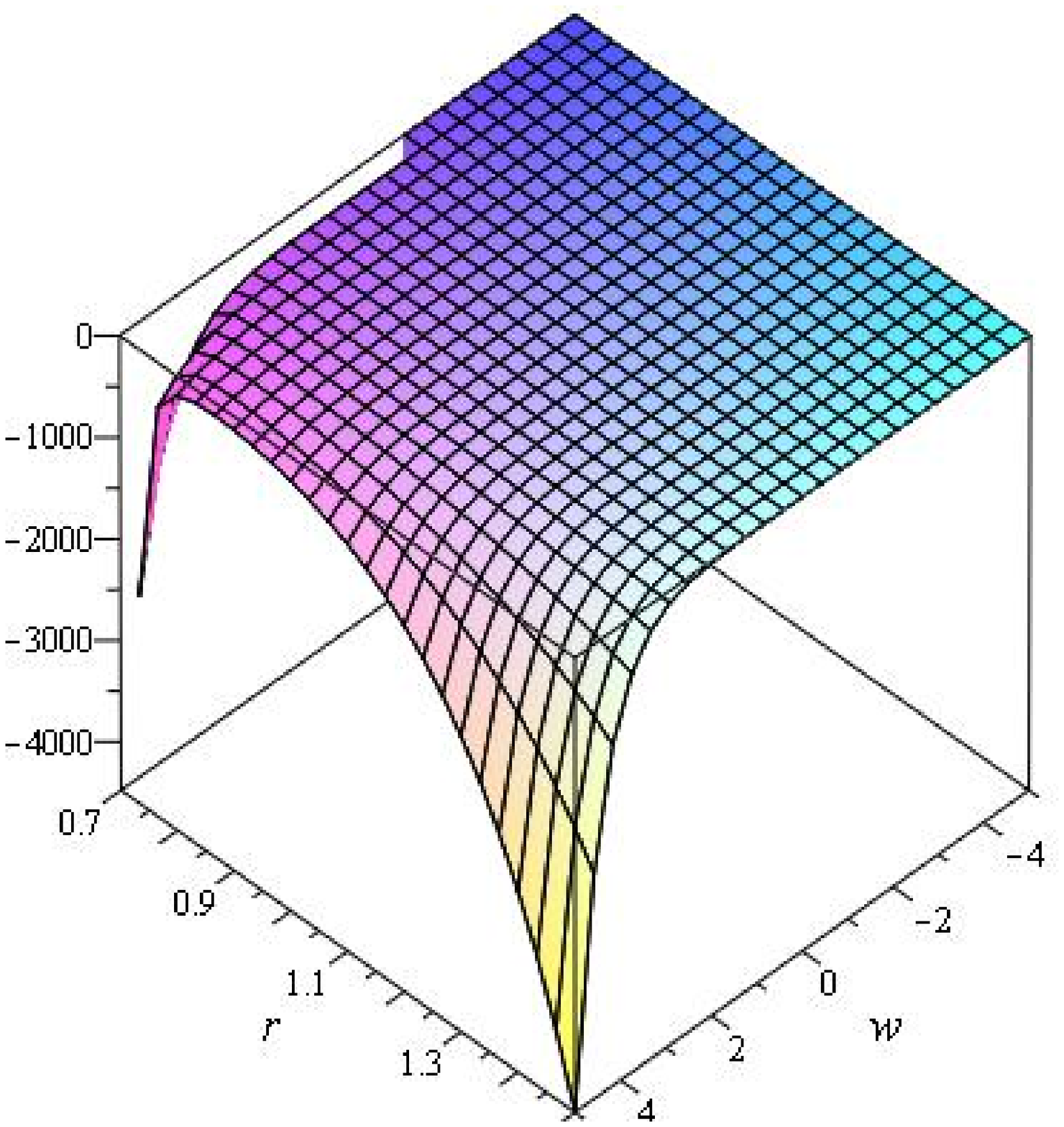}
		\centering FIG.10(E)
	\end{minipage}
	\begin{minipage}{.32\textwidth}
		\centering
		\includegraphics[width=.6\linewidth]{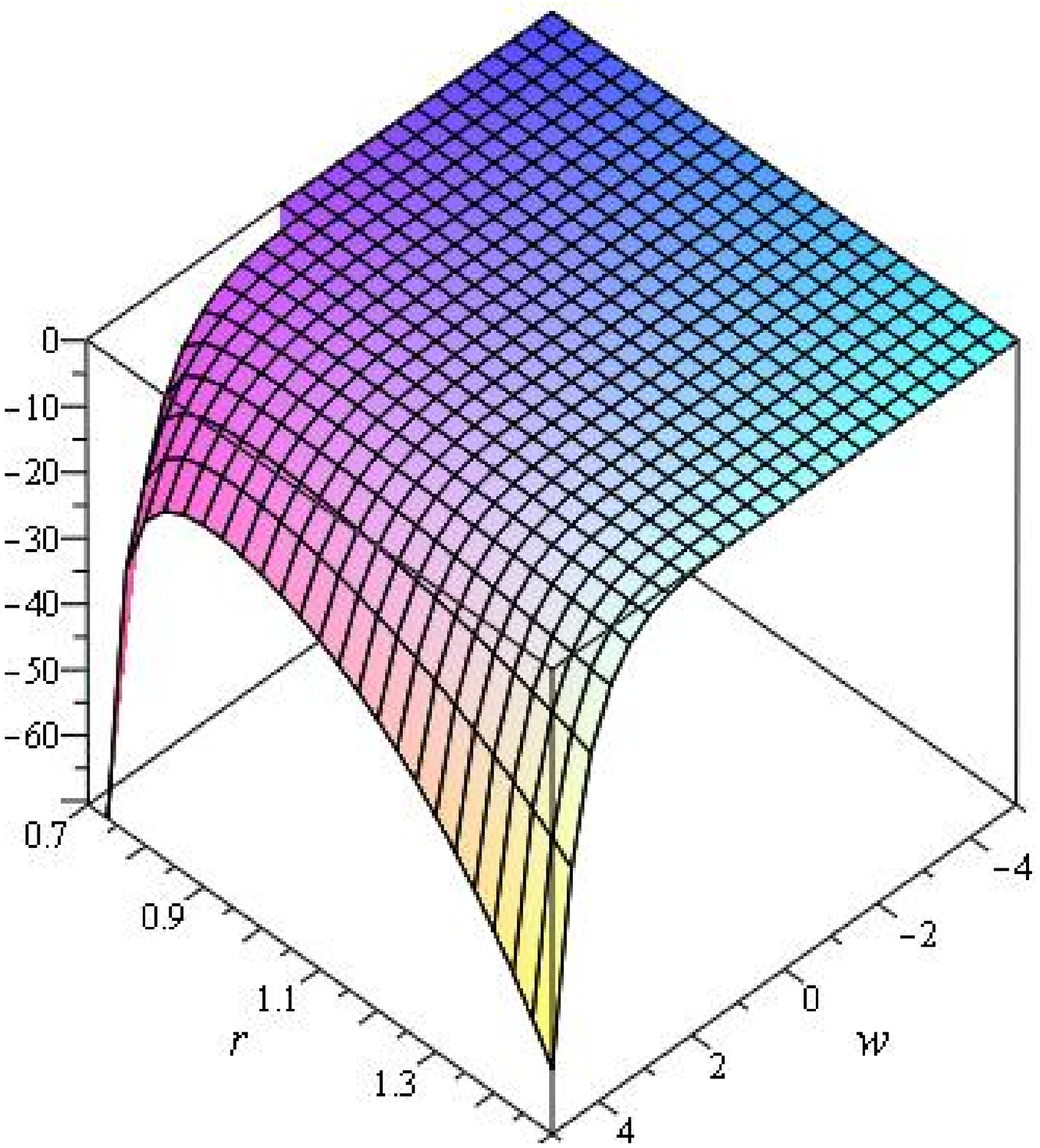}
		\centering FIG.10(F)
	\end{minipage}
		\caption{Behavior of $\rho$ (FIG.10(A)), $\rho+p_r$ (FIG.10(B)), $\rho+p_t$ (FIG.10(C)), $\rho+p_r+2p_t$ (FIG.10(D)), $\rho-|p_r|$ (FIG.10(E)) and $\rho-|p_t|$ (FIG.10(F)) diagrams have been plotted  against $r$ and $\omega$ for shape function(2) when $C_1=1$, $-5\leq\omega\leq5$ and $r_0=0.7$.}	\label{fig9.2}
\end{figure} 
 We have drawn the 3D figures (Figs.\ref{fig9.1} and \ref{fig9.2}) to observe the behavior of $\rho$, $\rho+p_r$, $\rho+p_t$, $\rho+p_r+2p_t$, $\rho-|p_r|$, $\rho-|p_t|$ in the anisotropic scenario for the shape functions (1) and (2), respectively.
\section{Isotropic matter field and wormhole solutions in $f(R)$-gravity theory}
\label{seciv}
For the isotropic fluid solutions, we consider $p_r=p_t=p$. Thus equations (\ref{eq14}) and (\ref{eq15}) lead to 
\begin{equation}\label{eq40}
F^{\prime\prime}(r)\left(1-\frac{b(r)}{r}\right)-F^{\prime}\left(\frac{1}{r}+\frac{b^{\prime}(r)}{2r}-\frac{3b(r)}{2r^2}\right)+F\left(\frac{3b(r)}{2r^3}-\frac{b^\prime}{2r^2}\right)=0.
\end{equation}
Now, for the same choice of shape function (1), $b(r)=\frac{r_0^n}{r^{n-1}}$ for some $n>0$, the above equation reduces to
\begin{equation}\label{eq41}
F^{\prime\prime}\left(\frac{1}{r_0^n}-\frac{1}{r^n}\right)-F^\prime\left(\frac{1}{r_0^nr}-\frac{n+2}{2r^{n+1}}\right)+F\frac{n+2}{2r^{n+2}}=0.
\end{equation}
From the equation(\ref{eq41}), for $n=2$ we obtain the solution
\begin{equation}\label{eq42}
F(r)=C_1P\left(\frac{1}{2},\frac{1}{2}\sqrt{17},\sqrt{\frac{-r^2+r_0^2}{r_0^2}}\right)r^{3/2}+C_2Q\left(\frac{1}{2},\frac{1}{2}\sqrt{17},\sqrt{\frac{-r^2+r_0^2}{r_0^2}}\right)r^{3/2},
\end{equation}
where $C_1$, $C_2$ are arbitrary constants and $P$, $Q$ are Legendre and associated Legendre functions, respectively. Now using equation (\ref{eq13}) we get the expression of energy density which is given by
\begin{equation}
\rho=-\frac{r_0^2}{r^4}\Biggl[C_1P\left(\frac{1}{2},\frac{1}{2}\sqrt{17},\sqrt{\frac{-r^2+r_0^2}{r_0^2}}\right)r^{3/2}+C_2Q\left(\frac{1}{2},\frac{1}{2}\sqrt{17},\sqrt{\frac{-r^2+r_0^2}{r_0^2}}\right)r^{3/2}\Biggr].
\end{equation}
 In this case, Ricci scalar is given by $R=-2\frac{r_0^2}{r^4}$ (using equation (\ref{eq11})) {\it i.e.} $r=\left(\frac{-2r_0^2}{R}\right)^{1/4}$, and hence at the throat we have $r_0=\left(\frac{-2}{R_0}\right)^{1/2}$. Putting these relations in equation (\ref{eq42}) provides the form of F(R), which is given by
 \begin{eqnarray}
 F(R)&=&C_1P\left(\frac{1}{2},\frac{1}{2}\sqrt{17},\sqrt{-\left(\frac{R_0}{R}\right)^{1/2}+1}\right)\left(\frac{4}{RR_0}\right)^{3/8}\nonumber\\
 &~&+~C_2Q\left(\frac{1}{2},\frac{1}{2}\sqrt{17},\sqrt{-\left(\frac{R_0}{R}\right)^{1/2}+1}\right)\left(\frac{4}{RR_0}\right)^{3/8}.
 \end{eqnarray}

 Now we can get the specific form of $f(R)$ from the given integral (existence of the integration is assured from the continuity of the integrand)
 \begin{equation}
 f(R)=\int_{R_0}^{R}F(R)dR.
 \end{equation}
 
 If we consider the same choice of shape function (2), $b(r)=\frac{r}{1+r-r_0}$ , $0<r_0<1$, the above equation (\ref{eq40}) reduces to
 \begin{equation}\label{eq46}
 F^{\prime\prime}2r^2\left(r-r_0\right)(1+r-r_0)-F^\prime r(2r^2+2r_0^2+r-2r_0-4rr_0)+F(2+3r-2r_0)=0.
 \end{equation}
 From the equation (\ref{eq46}), we obtain the solution
 \begin{eqnarray}\label{eq55}
 F(r)&=&C_1r^{\frac{\sqrt{r_0(r_0+1)}+r_0}{r_0}}G\bigg(\frac{-1+r_0}{r_0}, \frac{1}{2}\frac{8(r_0(r_0+1))^3/2+(-9r_0-8r_0^2)\sqrt{r_0(r_0+1)}-2r_0}{r_0^3},\nonumber\\ &~&\frac{\sqrt{r_0(r_0+1)}-r_0}{r_0},\frac{\sqrt{r_0(r_0+1)}+r_0}{r_0},\frac{r_0+2\sqrt{r_0(r_0+1)}}{r_0},\frac{1}{2}, \frac{r}{r_0}\bigg)\nonumber\\
 &~&+~C_2r^{\frac{-\sqrt{r_0(r_0+1)}+r_0}{r_0}}G\bigg(\frac{-1+r_0}{r_0}, \frac{1}{2}\frac{8(r_0(r_0+1))^3/2+(-7r_0-8r_0^2)\sqrt{r_0(r_0+1)}-2r_0}{r_0^3},\nonumber\\
 &~& \frac{-\sqrt{r_0(r_0+1)}+r_0}{r_0}, \frac{-\sqrt{r_0(r_0+1)}-r_0}{r_0},\frac{r_0-2\sqrt{r_0(r_0+1)}}{r_0},\frac{1}{2}, \frac{r}{r_0}\bigg),
 \end{eqnarray}
 where $C_1$, $C_2$ are arbitrary constants and $G$ is the Heun function.
  Note that for the third choice of the shape function it is not possible to find $F(r)$ explicitly and hence it is not presented here.

  \section{Embedding diagrams of wormholes}
  \label{secv}
  
\begin{figure}[h]
  	\centering
  	\begin{minipage}{.5\textwidth}
  		\centering
  		\includegraphics[width=.6\linewidth]{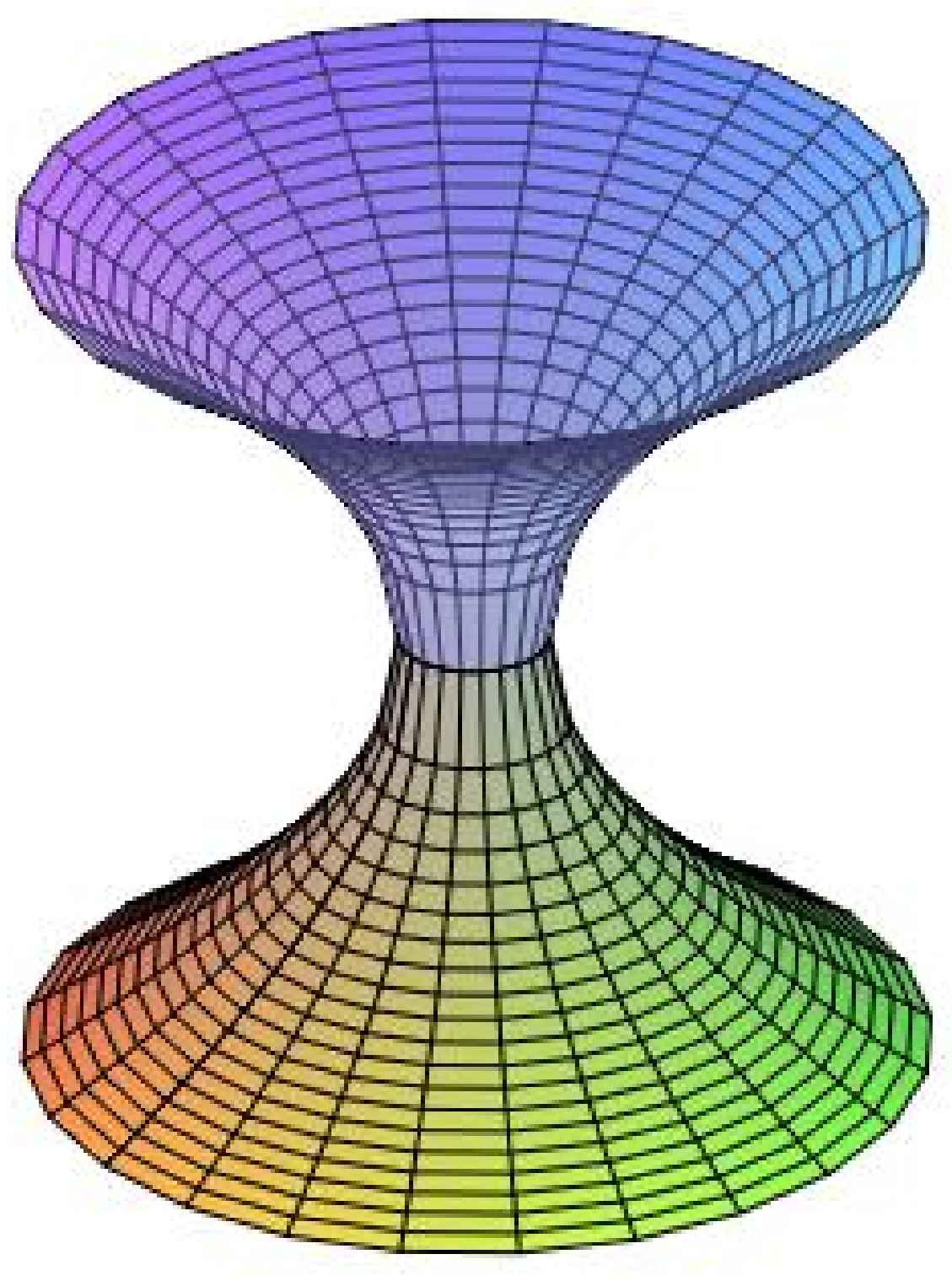}
  		\centering FIG.11(A)	
  	\end{minipage}
  	\begin{minipage}{.5\textwidth}
  		\centering
  		\includegraphics[width=.6\linewidth]{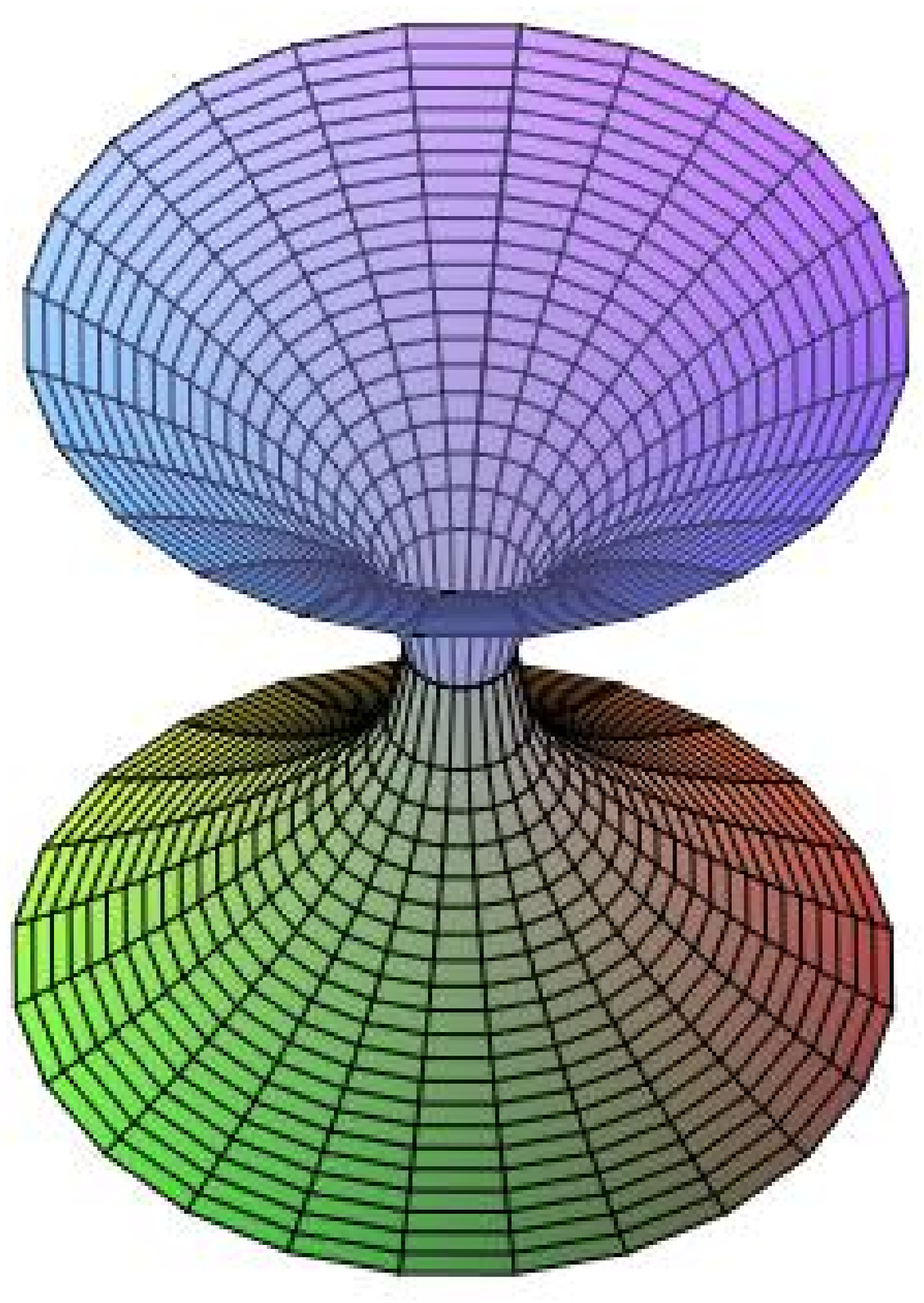}
  		\centering FIG.11(B)	
  	\end{minipage}
  	
  	\caption{ Embedding diagram for the shape function 1 (FIG.11(A)), and for shape function 2 (FIG.11(B)) considering the numerical values $n=0.9$, $r_0=0.5$ and $r_0=0.5$ for these shape functions, respectively.}
  	\label{fig9}
  \end{figure}

\begin{figure}[!htb]
	\centering
	\includegraphics[width=.4\linewidth]{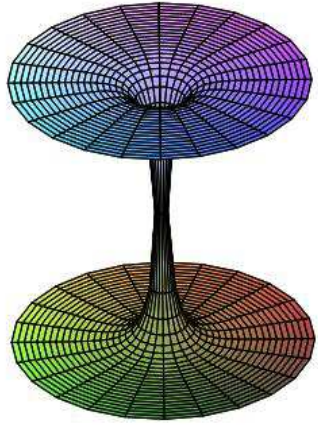}
	\caption{ Embedding diagram for the shape function 3 for  $r_0=0.5$.}
	\label{fig10}
\end{figure}

  To study the wormhole topology, let us consider the $2D$ hyper surface $\sum$: $t$=constant, $\theta=\pi/2$. The geometry is characterized by\cite{r4},\cite{r20} 
  \begin{equation} \label{eq36.2} d{S^2_{\sum}}=\frac{dr^2}{1-\frac{b(r)}{r}}+r^2d\phi^2.
  \end{equation}
  One may consider this $2D$ hyper surface $\sum$ embedded as rotational surface $ z=z(r,\phi)$ into the Euclidean space with metric \begin{equation} \label{eq36.3} d{S^2_{\sum}}=\left[1+\left(\frac{dz}{dr}\right)^2\right]dr^2+r^2d\phi^2 ,\end{equation} in cylindrical co-ordinates $(r,\phi,z)$. Thus, comparing (\ref{eq36.2}) and (\ref{eq36.3}), one may obtain the expression for embedding function as 
  \begin{equation} \label{eq27.1} z(r)=\int_{r_0}^{r}\sqrt{\frac{(b/r)}{1-(b/r)}}dr.
  \end{equation}
  Embedding diagrams of wormhole are drawn in FIG.\ref{fig9}--FIG.\ref{fig10}, considering the mentioned shape functions.
 \section{summary and concluding remarks}
 \label{secvi}
 This work addresses the following two questions in connection to the formation of traversable wormhole. The first question is that whether the general mechanism developed recently \cite{r15} for constructing wormhole solution in Einstein gravity can be extended to $f(R)$-modified gravity theory. The second question is whether the claim in recent past \cite{r17} that formation of wormhole with isotropic matter source in Einstein gravity needs non-zero tidal force is true or not in the present $f(R)$-modified gravity theory. Here, three different types of power-law form have been chosen for the shape function and wormhole solutions has been evaluated for anisotropic fluid using the general technique proposed in \cite{r15}. It is found that all the energy conditions has been satisfied for the first choice of the shape function in some suitable region of the domain space (in fig.1 to fig.6). For the second choice of the shape function although the energy density is positive but still no energy condition is satisfied (in fig.7 and fig.8). For the third choice of $b(r)$ due to complicated expressions of different energy conditions, it is not possible to infer about their validity. Hence the general technique adopted in \cite{r15} can be extended to any modified gravity theory for formation of wormholes.\\
 To address the second question, wormhole solutions for isotropic matter have been constructed for the first two choices of the shape function and vanishing redshift function ({\it i.e.} $\phi(r)=0$). Therefore, from the above study, one may conclude that wormhole solutions with isotropic matter field and zero tidal force are possible in modified gravity theory in contrast to Einstein gravity. Further, this new technique for determination of wormhole solution (valid for any gravity theory) helps us to examine whether a typical wormhole geometry is possible in a particular gravity theory or not. Finally, for future work, it would be interesting to examine that such technique of determination of wormhole solution is possible for evolving wormhole geometry.
 \section*{Acknowledgement}
 The author B.G.  is thankful to UGC for NET-JRF (F.No. 16-9(June 2019)/2019(NET/CSIR)) and the Department of
 Mathematics, Jadavpur University where a part of the work was completed. S.C. thanks Science and Engineering Research Board (SERB) for awarding
 MATRICS Research Grant support (File No. MTR/2017/000407) .

\end{document}